\pdfoutput=1
\RequirePackage{ifpdf}
\ifpdf 
\documentclass[pdftex]{sigma}
\else
\documentclass{sigma}
\fi

\numberwithin{equation}{section}

\newtheorem{Theorem}{Theorem}[section]
\newtheorem{Proposition}[Theorem]{Proposition}
 { \theoremstyle{definition}

 }

\usepackage{esint}

\usepackage{mathtools}
\mathtoolsset{showonlyrefs}

\usepackage{tikz}
\usetikzlibrary{
	arrows,
	decorations.pathreplacing,
	decorations.markings
}

\let\leq=\leqslant
\let\geq=\geqslant

\newcommand{\bra}[1]{\langle #1 \rvert}
\newcommand{\ket}[1]{\lvert #1 \rangle}

\def\Hcal{\mathcal{H}}

\def\Vcal{\mathcal{V}}

\def\Cbb{\mathbb{C}}

\def\ua{\uparrow}
\def\da{\downarrow}
\def\kua{\ket{\uparrow}}
\def\kda{\ket{\downarrow}}
\def\Ua{\Uparrow}
\def\Da{\Downarrow}

\newcommand\wt{\widetilde}

\DeclareMathOperator*{\sgn}{\mathrm{sgn}}

\DeclareMathOperator{\erf}{\mathrm{erf}}
\DeclareMathOperator{\erfc}{\mathrm{erfc}}

\newcommand{\rme}{\mathrm{e}}
\newcommand{\rmi}{\mathrm{i}}
\newcommand{\rmd}{\mathrm{d}}

\newcommand{\JP}[4]{{}\,P_{#1}^{(#2,#3)}(#4)}

\newcommand{\Ftwoone}[4]{%
\,{}_{2}F_{1}\bigg(\genfrac{}{}{0pt}{}{#1,\, #2}{#3} \bigg\vert #4\bigg)
}

\newcommand{\FtwoonePrime}[4]{%
\,{{}_{2}F_{1}}'\bigg(\genfrac{}{}{0pt}{}{#1,\, #2}{#3} \bigg\vert #4\bigg)
}

\begin{document}
\allowdisplaybreaks

\renewcommand{\thefootnote}{}

\newcommand{\arXivNumber}{2108.06190}

\renewcommand{\PaperNumber}{111}

\FirstPageHeading

\ShortArticleName{Boundary One-Point Function of the Rational Six-Vertex Model}

\ArticleName{Boundary One-Point Function of the Rational\\ Six-Vertex Model with Partial Domain Wall Boundary\\ Conditions:
Explicit Formulas and Scaling Properties\footnote{This paper is a~contribution to the Special Issue on Mathematics of Integrable Systems: Classical and Quantum in honor of Leon Takhtajan.

~~\,The full collection is available at \href{https://www.emis.de/journals/SIGMA/Takhtajan.html}{https://www.emis.de/journals/SIGMA/Takhtajan.html}}}

\Author{Mikhail D.~MININ and Andrei G.~PRONKO}
\AuthorNameForHeading{M.D.~Minin and A.G.~Pronko}
\Address{Steklov Mathematical Institute, Fontanka 27, St. Petersburg, 191023, Russia}
\Email{\href{mailto:mininmd96@gmail.com}{mininmd96@gmail.com}, \href{mailto:agp@pdmi.ras.ru}{agp@pdmi.ras.ru}}

\ArticleDates{Received August 16, 2021, in final form December 18, 2021; Published online December 25, 2021}

\Abstract{We consider the six-vertex model with the rational weights on an $s\times N$ square lattice, $s\leq N$, with partial domain wall boundary conditions. We study the one-point function at the boundary where the free boundary conditions are imposed. For a finite lattice, it can be computed by the quantum inverse scattering method in terms of determinants. In the large $N$ limit, the result boils down to an explicit terminating series in the parameter of the weights. Using the saddle-point method for an equivalent integral representation, we show that as $s$ next tends to infinity, the one-point function demonstrates a step-wise behavior; at the vicinity of the step it scales as the error function. We also show that the asymptotic expansion of the one-point function can be computed from a second-order ordinary differential equation.}


\Classification{05A19; 05E05; 82B23}

\begin{flushright}
\begin{minipage}{86mm}\it
Dedicated to Professor Leon Armenovich Takhtajan\\
on the occasion of his 70$\,{}^{th}$ birthday
\end{minipage}
\end{flushright}

\renewcommand{\thefootnote}{\arabic{footnote}}
\setcounter{footnote}{0}

\section{Introduction}

Integrability of 1D quantum systems is closely related to the notion of the
Bethe anzats and, even more significantly, to its algebraic version,
introduced by Takhtajan and Faddeev \cite{TF-79}. In a~more general sense, the
proposed method consists in quantization of the monodromy matrix of a classical
integrable system in the settings of an inverse scattering problem,~-- hence
the name quantum inverse scattering method (QISM)~\cite{KBI-93}.

The central role in the QISM is played by commutation relations between
operators being matrix elements of the quantum monodromy matrix. These
commutation relations are described by the so-called $R$-matrix, which also has
the meaning of a matrix of Boltzmann weights of some lattice model of 2D
statistical mechanics. The crucial property of the $R$-matrix is that it obeys the
Yang--Baxter relation, which leads to integrability of the underlying lattice
model. This means that its transfer matrix in the case of periodic boundary
conditions appears to be a~generating function of an infinite number of commuting
operators.

A famous and important example is provided by the six-vertex model,
also known as the ice-type model~\cite{B-82}. It is related to the so-called
trigonometric $R$-matrix, and, as a special case, the rational $R$-matrix.
These two $R$-matrices describe commutation relations for the
elements of the monodromy matrices of the Heisenberg $XXZ$ and $XXX$ spin chains,
respectively.

In \cite{K-82}, Korepin, in addressing the proof within the QISM of the Gaudin
hypothesis on the norm of Bethe states, showed that the scalar products
in the $N$-particle sector are expressed in terms of a special object, which
has a meaning as the partition function of the six-vertex model on a
finite-size domain constructed by intersection of $N$ vertical and $N$
horizontal lines, the so-called $N\times N$ lattice, with special fixed
boundary conditions called domain wall boundary conditions (DWBC). In the
spin language for local states, typical for the QISM, they mean that the
states on each of the four boundaries are all the same, and have the opposite
values, spin up versus spin down, at the opposite boundaries of the $N\times
N$ lattice, so as a result, the boundary spins around the $N\times N$ lattice
form a domain wall.

\looseness=1 A milestone result was later obtained by Izergin \cite{I-87}, who wrote down
an explicit determinant formula satisfying all the properties listed by
Korepin in \cite{K-82}, which fix completely the partition function~\cite{ICK-92}.
Besides applications to study of correlation functions of the
related 1D quantum systems \cite{KKMST-09,KMST-02,KBI-93},
this result has turned out to be of
importance, e.g., in combinatorics, for proving long-standing conjectures on
enumerations of alternating-sign matrices \cite{Ku-96,Ze-96}. Further study
of the six-vertex model with DWBC has showed that it provides an example of
a statistical mechanics model where the boundary conditions significantly
affect bulk thermodynamic quantities \cite{BL-14, KZj-00}. That is closely related
to presence of nontrivial limit shapes in the scaling limit, as has been
established both numerically and
analytically \cite{AR-05,BR-20,CP-09,CP-08,KP-20,LKRV-18,SZ-04}.

The six-vertex model is also very interesting in the case where DWBC are relaxed at
one of their four boundaries. These are the so-called partial
domain wall boundary conditions (pDWBC). Specifically, the six-vertex model is
considered on an $s\times N$ lattice, $s\leq N$, and pDWBC mean that DWBC are kept
at the two boundaries of length $s$ and one boundary of length $N$, while the
remaining boundary of length $N$ is left free, that is, the summation
over all possible states is performed. This model arises, in particular, in the
context of calculations of correlation function
in $\mathcal{N}=4$ supersymmetric Yang--Mills theory
\cite{EGSV-11a,EGSV-11b,GSV-12,JKKS-16}.

Unlike the case of DWBC, in the case of pDWBC determinant representations are
only known for the rational weights and non-symmetric trigonometric weights
satisfying the Gwa--Spohn stochasticity condition \cite{GS-92}. Determinant
representations have been obtained by Foda and Wheeler
\cite{F-12,FW-12, W-11} and Kostov~\cite{K-12a,K-12b}; thermodynamics has been
studied by Bleher and Liechty~\cite{BL-15}. For arbitrary symmetric weights,
apart from the free-fermion case, it seems no general determinant formula
for the partition function exists, although in the limit $N\to\infty$ the
partition function can be given in terms of a Pfaffian~\cite{PP-19}.

In the present paper, limiting ourselves by the case of
rational weights, we study the
simplest one-point correlation function describing polarization at the
boundary where the free conditions are imposed.
Our central result is an explicit formula for this function
in the limit $N\to \infty$ (see Proposition \ref{Prop-main-result} in Section~\ref{Sec31}).
In this limit the lattice turn into the semi-infinite lattice strip;
for simplicity, we call it semi-infinite lattice.
We obtain this formula by two methods. We first show that it can be
conjectured using a formulation of the model in terms off-shell Bethe states
in the coordinate form, and next we give a proof using the QISM formalism.
In the latter, a~contour integral formula for the one-point function appears,
which turns out to be convenient for proving various other representations.

We also study the scaling properties of the boundary one-point
function as $s$ tends to infinity and the mesh size of the lattice
vanishes, so that lattice is scaled to a semi-infinite strip of a unit width.
We find that the one-point function demonstrates a step-wise behavior; at the
vicinity of the step it scales as a (complementary) error function.
Again, we use two methods to obtain the result. One is the saddle-point method
applied to the contour integral formula. Since it appears rather subtle
in its topological part, we show that the same result can also be obtained from
an ordinary differential equation (ODE) for the one-point function
by means of the method proposed in \cite{KP-16}.
This method allows for construction
of the asymptotic expansion recursively starting from the leading term, which
is simply determined in our case as a proper root of an algebraic cubic equation.

The present paper can be seen as a complementary to \cite{MP-19} where
we have derived a formula in terms of a sum over the
Jacobi polynomials. The main difference is that in our QISM calculations here
we put forward the determinant formula by Kostov. This makes it possible
to take the $N\to \infty$ limit before the homogeneous limit,
that significantly simplifies subsequent calculations.

The paper is organized as follows. In Section~\ref{section2} we give the definition of the
model and recall the determinant formulas for the partition function. Section~\ref{section3}
is devoted to the boundary one-point function and the QISM calculations.
Section~\ref{section4} contains derivation of the step-function behavior of the one-point
function in the scaling limit; we also prove that it obeys a second-order ODE
and explain how this equation can be used to build the asymptotic expansion. In
Section~\ref{section5} we briefly discuss our results, also in application
to the phase separation phenomena.

\section[Six-vertex model with partial domain wall boundary conditions]{Six-vertex model with partial domain wall boundary\\ conditions}\label{section2}

In the section we define the model and give some basic results, such as
determinant formulas for the partition function in the case of
rational Boltzmann weights. We also explain how it simplifies in the
limit of a semi-infinite lattice.

\subsection{Definition of the model}\label{Sec21}

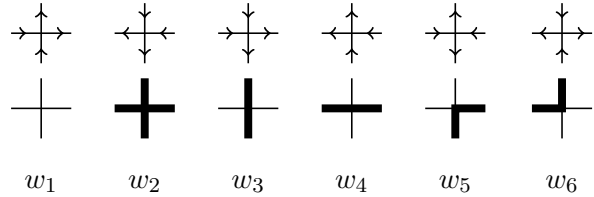
\begin{figure}\centering
\begin{tikzpicture}[scale=.5]
\draw [semithick] (0.2,3)--(1.8,3);
\draw [semithick] (1,2.2)--(1,3.8);
\draw [thick] [->] (0.5,3)--(.6,3);
\draw [thick] [->] (1.5,3)--(1.6,3);
\draw [thick] [->] (1,2.5)--(1,2.6);
\draw [thick] [->] (1,3.5)--(1,3.6);
\draw [semithick] (0.2,1)--(1.8,1);
\draw [semithick] (1,0.2)--(1,1.8);
\node at (1,-1) {$w_1$};
\end{tikzpicture}
\quad
\begin{tikzpicture}[scale=.5]
\draw [semithick] (0.2,3)--(1.8,3);
\draw [semithick] (1,2.2)--(1,3.8);
\draw [thick] [->] (0.5,3)--(0.4,3);
\draw [thick] [->] (1.5,3)--(1.4,3);
\draw [thick] [->] (1,2.5)--(1,2.4);
\draw [thick] [->] (1,3.5)--(1,3.4);
\draw [line width=3] (0.2,1)--(1.8,1);
\draw [line width=3] (1,0.2)--(1,1.8);
\node at (1,-1) {$w_2$};
\end{tikzpicture}
\quad
\begin{tikzpicture}[scale=.5]
\draw [semithick] (0.2,3)--(1.8,3);
\draw [semithick] (1,2.2)--(1,3.8);
\draw [thick] [->] (0.5,3)--(0.6,3);
\draw [thick] [->] (1.5,3)--(1.6,3);
\draw [thick] [->] (1,2.5)--(1,2.4);
\draw [thick] [->] (1,3.5)--(1,3.4);
\draw [semithick] (0.2,1)--(1.8,1);
\draw [line width=3] (1,0.2)--(1,1.8);
\node at (1,-1) {$w_3$};
\end{tikzpicture}
\quad
\begin{tikzpicture}[scale=.5]
\draw [semithick] (0.2,3)--(1.8,3);
\draw [semithick] (1,2.2)--(1,3.8);
\draw [thick] [->] (0.5,3)--(0.4,3);
\draw [thick] [->] (1.5,3)--(1.4,3);
\draw [thick] [->] (1,2.5)--(1,2.6);
\draw [thick] [->] (1,3.5)--(1,3.6);
\draw [line width=3] (0.2,1)--(1.8,1);
\draw [semithick] (1,0.2)--(1,1.8);
\node at (1,-1) {$w_4$};
\end{tikzpicture}
\quad
\begin{tikzpicture}[scale=.5]
\draw [semithick] (0.2,3)--(1.8,3);
\draw [semithick] (1,2.2)--(1,3.8);
\draw [thick] [->] (0.5,3)--(.6,3);
\draw [thick] [->] (1.5,3)--(1.4,3);
\draw [thick] [->] (1,2.5)--(1,2.4);
\draw [thick] [->] (1,3.5)--(1,3.6);
\draw [semithick] (0.2,1)--(1,1)--(1,1.8);
\draw [line width=3] (1,0.2)--(1,1)--(1.8,1);
\node at (1,-1) {$w_5$};
\end{tikzpicture}
\quad
\begin{tikzpicture}[scale=.5]
\draw [semithick] (0.2,3)--(1.8,3);
\draw [semithick] (1,2.2)--(1,3.8);
\draw [thick] [->] (0.5,3)--(.4,3);
\draw [thick] [->] (1.5,3)--(1.6,3);
\draw [thick] [->] (1,2.5)--(1,2.6);
\draw [thick] [->] (1,3.5)--(1,3.4);
\draw [line width=3] (0.2,1)--(1,1)--(1,1.8);
\draw [semithick] (1,0.2)--(1,1)--(1.8,1);
\node at (1,-1) {$w_6$};
\end{tikzpicture}
\caption{Admissible vertex configurations in terms of arrows (first row),
solid lines (second row), and their Boltzmann weights $w_i$ (third row).}\label{fig-SixVertices}
\end{figure}

We consider the six-vertex model on an $s\times N$ square lattice,
$s\leq N$, i.e., a lattice obtained by intersection of $s$ horizontal
and $N$ vertical lines.
The configurations of the six-vertex model are usually defined
in terms of arrows placed on edges, or, equivalently, in terms of solid lines.
The correspondence between arrows and lines is the following:
if the edge has left or down arrow,
then it carries a solid line, otherwise, if it has
right or up arrow, it is empty. The allowed arrow configurations around a vertex
are only those which contain an equal number of incoming and outgoing arrows, or,
in terms of lines, only those which conserve the number of lines passing along
the NE-SW direction, see Figure~\ref{fig-SixVertices}, where the six
allowed vertex configurations are given in the standard order~\cite{B-82}.
The partition function of the six-vertex model is defined as
\begin{gather}\label{sumover}
Z= \sum_{\mathcal{C}\in\Omega} \prod_{i=1,\dots, 6} w_i^{n_i(\mathcal{C})},
\end{gather}
where $\Omega$ is the set of allowed arrow configurations,
$n_i(\mathcal{C})$ is the number of vertices of type $i$ ($i=1,\dots,6$) in
the configuration $\mathcal{C}$, and $w_i$ is the corresponding Boltzmann
weight of the vertex.

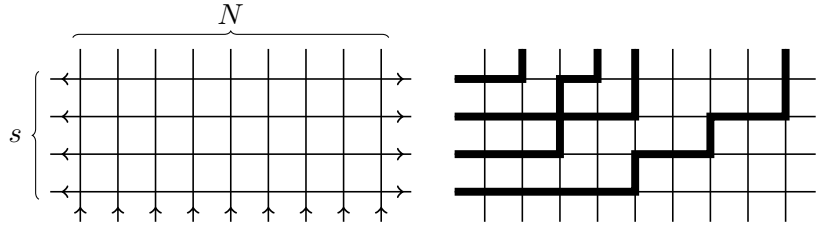
\begin{figure}\centering

\usetikzlibrary{decorations.pathreplacing}

\begin{tikzpicture}[scale=.5]
\draw [semithick] (0.2,1)--(9.8,1);
\draw [semithick] (0.2,2)--(9.8,2);
\draw [semithick] (0.2,3)--(9.8,3);
\draw [semithick] (0.2,4)--(9.8,4);
\draw [semithick] (1,0.2)--(1,4.8);
\draw [semithick] (2,0.2)--(2,4.8);
\draw [semithick] (3,0.2)--(3,4.8);
\draw [semithick] (4,0.2)--(4,4.8);
\draw [semithick] (5,0.2)--(5,4.8);
\draw [semithick] (6,0.2)--(6,4.8);
\draw [semithick] (7,0.2)--(7,4.8);
\draw [semithick] (8,0.2)--(8,4.8);
\draw [semithick] (9,0.2)--(9,4.8);
\draw [thick] [->] (.6,1)--(.5,1);
\draw [thick] [->] (.6,2)--(.5,2);
\draw [thick] [->] (.6,3)--(.5,3);
\draw [thick] [->] (.6,4)--(.5,4);
\draw [thick] [->] (9.5,1)--(9.6,1);
\draw [thick] [->] (9.5,2)--(9.6,2);
\draw [thick] [->] (9.5,3)--(9.6,3);
\draw [thick] [->] (9.5,4)--(9.6,4);
\draw [thick] [->] (1,.5)--(1,.6);
\draw [thick] [->] (2,.5)--(2,.6);
\draw [thick] [->] (3,.5)--(3,.6);
\draw [thick] [->] (4,.5)--(4,.6);
\draw [thick] [->] (5,.5)--(5,.6);
\draw [thick] [->] (6,.5)--(6,.6);
\draw [thick] [->] (7,.5)--(7,.6);
\draw [thick] [->] (8,.5)--(8,.6);
\draw [thick] [->] (9,.5)--(9,.6);
\draw [decorate,decoration={brace}]
(.8,5.1) -- (9.2,5.1) node [midway,yshift=9pt] {$N$};
\draw [decorate,decoration={brace}]
(-.1,0.8) -- (-.1,4.2) node [midway,xshift=-9pt] {$s$};
\end{tikzpicture}
\quad
\begin{tikzpicture}[scale=.5]
\draw [semithick] (0.2,1)--(9.8,1);
\draw [semithick] (0.2,2)--(9.8,2);
\draw [semithick] (0.2,3)--(9.8,3);
\draw [semithick] (0.2,4)--(9.8,4);
\draw [semithick] (1,0.2)--(1,4.8);
\draw [semithick] (2,0.2)--(2,4.8);
\draw [semithick] (3,0.2)--(3,4.8);
\draw [semithick] (4,0.2)--(4,4.8);
\draw [semithick] (5,0.2)--(5,4.8);
\draw [semithick] (6,0.2)--(6,4.8);
\draw [semithick] (7,0.2)--(7,4.8);
\draw [semithick] (8,0.2)--(8,4.8);
\draw [semithick] (9,0.2)--(9,4.8);
\draw [line width=3] (0.2,4)--(2,4)--(2,4.8);
\draw [line width=3] (0.2,3)--(3,3)--(3,4)--(4,4)--(4,4.8);
\draw [line width=3] (0.2,2)--(3,2)--(3,3)--(5,3)--(5,4.8);
\draw [line width=3] (0.2,1)--(5,1)--(5,2)--(7,2)--(7,3)--(9,3)--(9,4.8);
\end{tikzpicture}
\caption{Partial domain wall boundary conditions: the arrow states on external
edges on the left, bottom and right boundaries are fixed, while on the top
boundary the summation over all possible states is performed (left); one of the possible
configurations in terms of the solid lines (right).}
\label{fig-sxNlattice}
\end{figure}

The set $\Omega$ of allowed configurations
(i.e., over those the summation in~\eqref{sumover}
is performed) depends on the boundary conditions imposed. For our
$s\times N$ lattice (recall that $s\leq N$)
we take the so-called \emph{partial domain wall
boundary conditions} (pDWBC) which are defined as follows.
All arrows on the left and right boundaries are outgoing, those on
the bottom boundary are incoming, but on the top boundary the summation
over all possible arrow configurations is performed,
see Figure~\ref{fig-sxNlattice}.
In the special case $s=N$, the only possible configuration
on the top boundary is with all incoming arrows, that
corresponds to the domain wall boundary conditions~\cite{FW-12, K-82}.

Throughout this paper we take the Boltzmann weights invariant under reversal
of all arrows,
\begin{gather}\label{symmetric}
w_1 = w_2 =: a,\qquad
w_3 = w_4 =: b,\qquad
w_5 = w_6 =: c,	
\end{gather}
and, furthermore, of the form
\begin{gather}\label{weights}
a = 1,\qquad
b = t,\qquad
c = 1 - t,\qquad 0 \leq t < 1.
\end{gather}
The model with weights given by \eqref{symmetric} and \eqref{weights}
is known as the \emph{rational} six-vertex model, since it is
related to the rational $R$-matrix in the Yang--Baxter relation.
In terms of the parameter
\begin{gather*}
\Delta= \frac{a^2+b^2-c^2}{2ab},
\end{gather*}
relevant for the
description of the phase diagram of the symmetric six-vertex model (see, e.g.,~\cite{B-82}),
it corresponds to the case $\Delta=1$, where among the two possible
branches, $a<b$ and $a>b$, the latter has been chosen. This means that
the weights \eqref{weights} have the following
important property
\begin{gather}\label{bc=a}
b+c=a=1.
\end{gather}
Here, the second equality is simply the choice of normalization,
which is crucial for considering the semi-infinite lattice limit
in what follows.
The first equality in \eqref{bc=a}
implies stochasticity of the model \cite{GS-92}.
Namely, interpreting the Boltzmann weight as a probability
of passing a solid line
through the vertex (see Figure~\ref{fig-SixVertices}), one finds that the total
probability is conserved (and equals to $1$); the parameter $t$ in \eqref{weights}
gives the probability of passing through the vertex without
change of the direction.

\subsection{Determinant formulas for the partition function}\label{Sec22}

To apply the quantum inverse
scattering method (QISM) we consider the inhomogeneous version of the model
where the Boltzmann weights $a$, $b$, and $c$ are site-dependent. Namely, we
introduce two sets of parameters, each parameter associated with a line of
the lattice. Let parameters $\lambda_1, \dots , \lambda_s$ correspond to
the horizontal lines enumerated from top to bottom and $\nu_1, \dots, \nu_N$
correspond to the vertical lines enumerated from left to
right.\footnote{In \cite{MP-19}, we have used the reverse order for the
vertical lines; the reason to change the convention is motivated by
calculations in Section~\ref{Sec32}.} The weights of the
vertex being at the intersection of $j$-th horizontal and $k$-th vertical
lines are
\begin{gather}\label{1bc}
a_{j k} = 1, \qquad
b_{j k} = b (\lambda_j, \nu_k), \qquad
c_{j k} = c (\lambda_j, \nu_k),
\end{gather}
where
\begin{gather}\label{bc-lanu}
b (\lambda, \nu) = \frac{\lambda- \nu}{\lambda- \nu + 1}, \qquad
c (\lambda, \nu) = \frac{1}{\lambda- \nu +1}.
\end{gather}
The case of all the parameters equal to each other within each set,
$\lambda_1=\dots=\lambda_s = \lambda$ and $\nu_1=\dots=\nu_N = \nu$,
where, without loss of generality, we can set $\nu=0$, corresponds
to the homogeneous model. In what follows, we call this case
the homogeneous limit.
The case where $\nu_1,\dots,\nu_N = 0$, but all $\lambda_1,\dots,\lambda_s$
remain distinct, will be referred to as the partial homogeneous limit.

Basically, two determinant formulas are known for
the partition function of the inhomogeneous model.
One is in terms of $N\times N$ determinant and it
has been obtained by Foda and Wheeler \cite{FW-12}.
\begin{Proposition}[Foda and Wheeler]
The partition function of the six-vertex model with pDWBC can be
written in the form
\begin{gather}\label{Z_Foda}
Z =
\frac{
\prod_{j = 1}^{s} \prod_{k = 1}^{N}
(\lambda_j-\nu_k)}{
\prod_{1 \leq j < k\leq s}(\lambda_k - \lambda_j)
\prod_{1 \leq j < k \leq N}(\nu_j - \nu_k)}
\det\left[
\begin{cases}
\varphi(\lambda_i,\nu_j), & i\leq s
\\
\nu_j^{N-i}, & i>s
\end{cases}
\right]_{i,j=1,\dots,N},
\end{gather}
where
\begin{gather}\label{varphi}
\varphi(\lambda,\nu)
= \frac1{\left(\lambda - \nu + 1\right)
\left(\lambda - \nu\right) }.
\end{gather}
\end{Proposition}
Apparently, \eqref{Z_Foda} provides a
generalization of the well-known formula \cite{I-87,ICK-92, K-82} for
the partition function of the six-vertex model in the case
of $s \times s$ lattice with DWBC
to the case of $s \times N$ lattice with pDWBC, valid for
the rational weights.\footnote{For the rational case, exactly which we discuss here,
this formula can also be found in the monograph by Gaudin~\cite[Proposition~J$ {}_2$]{G-83}, where it appears as a
representation for the off-shell Bethe
state with the consecutive values of the particles'
coordinates. However, to identify this quantity
as the partition function of the six-vertex model with
DWBC, some additional tools are needed;
this becomes apparent within the QISM formalism
(see discussion in Section~\ref{Sec31}).}
On the other hand, \eqref{Z_Foda} can be obtained from the DWBC case
but for the $N\times N$ lattice,
in the limit where $(N-s)$ parameters associated with the rows are sent to
infinity, $\lambda_{N-s+1},\dots,\lambda_N\to\infty$~\cite{FW-12}.

Another formula for the partition function is in terms of $s\times s$ determinant
and it is due to Kostov \cite{K-12a,K-12b}.
\begin{Proposition}[Kostov]
The partition function of the rational six-vertex model with pDWBC
admits the representation
\begin{gather}\label{Z_Kostov}
Z=\prod_{1\leq j<k\leq s} \frac{1}{\lambda_k-\lambda_j}
\det \left[\lambda_i^{j-1}-(\lambda_i+1)^{j-1}
\prod_{k=1}^N
b(\lambda_i,\nu_k)\right]_{i,j=1,\dots,s},
\end{gather}
where the function $b(\lambda,\nu)$ is defined in \eqref{bc-lanu}.
\end{Proposition}
The equivalence of \eqref{Z_Foda} and \eqref{Z_Kostov} was proved in~\cite{FW-12}.

Let us now discuss these representations in the homogeneous limit. We will
focus mostly on~\eqref{Z_Foda}, since these two representations at some stage
coincide explicitly, see the discussion below after~\eqref{ZPP}.

We first consider the partition
function in the partial homogeneous limit, where $\nu_1,\dots,\nu_N\allowbreak \to 0$,
but $\lambda_1,\dots,\lambda_s$ are still arbitrary. In this limit
\eqref{Z_Kostov} is non-singular, and can be trivially computed, while
treatment of \eqref{Z_Foda} requires some work. The key relation here is
that if $f_1(\nu),\dots,f_N(\nu)$ are at least $N-1$ times differentiable
functions with respect to their arguments, then
\begin{gather*}
\lim_{\nu_1,\dots,\nu_N\to \nu}
\prod_{1 \leq j < k \leq N} \frac{1}{\nu_k - \nu_j}
\det\left[f_i(\nu_j)\right]_{i,j=1,\dots,N}
=\frac{1}{\prod_{k=1}^{N-1} k!}
\det\big[f_i^{(j-1)}(\nu)\big]_{i,j=1,\dots,N}.
\end{gather*}
Using the property $\partial_\nu\varphi(\lambda,\nu)=-\partial_\lambda\varphi(\lambda,\nu)$, see \eqref{varphi},
and denoting $\varphi(\lambda)\equiv \varphi(\lambda,0)$,
we get
\begin{gather*}
\lim_{\nu_1,\dots,\nu_N\to 0}
\prod_{1 \leq j < k \leq N} \frac{1}{\nu_j - \nu_k}
\det\left[
\begin{cases}
\varphi(\lambda_i,\nu_j), & i\leq s \\
\nu_j^{N-i}, & i>s
\end{cases}
\right]_{i,j=1,\dots,N}
\\
\qquad{} =\frac{(-1)^{\frac{N(N-1)}{2}}}{\prod_{k=1}^{N-1}k!}
\det\left[
\begin{cases}
(-1)^{j-1}\varphi^{(j-1)}(\lambda_i), & i\leq s\\
(j-1)!\,\delta_{N-i,j-1}, & i>s
\end{cases}
\right]_{i,j=1,\dots,N}
\\
\qquad{}=\frac{(-1)^{s(N-s)}}{\prod_{k=1}^{s}(N-k)!}
\det\big[
\varphi^{(N-s+j-1)}(\lambda_i)
\big]_{i,j=1,\dots,s}.
\end{gather*}
Note that here the last determinant is of size $s\times s$.
As a result, for the partition function of the partially inhomogeneous model
we have
\begin{gather}\label{Zparthom}
Z= \frac{(-1)^{s(N-s)}}{\prod_{k=1}^{s}(N-k)!}
\frac{\prod_{j=1}^s\lambda_j^N}
{\prod_{1\leq j<k\leq s}(\lambda_k-\lambda_j)}
\det\big[\varphi^{(N-s+i-1)}(\lambda_j)\big]_{i,j=1,\dots,s}.
\end{gather}

Let us now consider the homogeneous limit of~\eqref{Z_Foda} in the full set of parameters,
namely, we also take the limit $\lambda_1,\dots,\lambda_s\to \lambda$
in addition to $\nu_1,\dots,\nu_N\to 0$.
Along the same lines, the following relation holds
\begin{gather*}
\lim_{\substack{
\nu_1,\dots,\nu_N\to 0\\
\lambda_1,\dots,\lambda_s\to \lambda}}
\prod_{1 \leq j < k \leq s}
\frac{1}{\lambda_k - \lambda_j}
\prod_{1 \leq j < k \leq N}
\frac{1}{\nu_j - \nu_k}\;
\det\left[
\begin{cases}
\varphi(\lambda_i,\nu_j), & i\leq s \\
\nu_j^{N-i}, & i>s
\end{cases}
\right]_{i,j=1,\dots,N}
\\
\qquad{} =\frac{(-1)^{s(N-s)}}{\prod_{k = 1}^{s} (N-k)! \prod_{k = 1}^{s-1} k!}
\det\big[
\varphi^{(N-s+i+j-2)}(\lambda)
\big]_{i,j=1,\dots,s}.
\end{gather*}
Hence, for the homogeneous model the following
formula is valid:
\begin{gather}\label{Zfullhom}
Z= \frac{(-1)^{s(N-s)}
\lambda^{sN}}{\prod_{k=1}^{s}(N-k)!\prod_{k=1}^{s-1}k!}
\det\big[
\varphi^{(N-s+i+j-2)}(\lambda)
\big]_{i,j=1,\dots,s}.
\end{gather}
Note that the result is given in terms of the determinant of a Hankel matrix,
similarly to the DWBC case \cite{BL-15, ICK-92}.

Now we consider how these representations can be made more explicit, by
computing derivatives of the function $\varphi(\lambda)$. Indeed,
let us switch from the variable $\lambda$ to the variable $t$
appearing in \eqref{weights},
\begin{gather}\label{tbla}
t = b(\lambda, 0) =
\frac{\lambda}{\lambda + 1},\qquad
\lambda=\frac{t}{1-t}.
\end{gather}
Writing the function $\varphi(\lambda)$ in the form
\begin{gather*}
\varphi (\lambda) = \frac{1}{\lambda}-\frac{1}{\lambda+1}
=\frac{1 - t}t - (1-t),
\end{gather*}
the $n$-th derivative with respect to $\lambda$ in terms of $t$ can be readily computed to be
\begin{gather}\label{varphidir}
\varphi^{(n)} (\lambda) = (-1)^n n!
\left(\frac{1 - t}{t}\right)^{n + 1} \big(1 - t^{n + 1}\big).
\end{gather}

Consider the partially inhomogeneous model with the weights
parameterized in terms of the parameters $t_j=b(\lambda_j,0)$, $j=1,\dots,s$.
Substituting \eqref{varphidir} into \eqref{Zparthom}, and taking into account that
\begin{gather}\label{lateprods}
\prod_{1\leq j<k\leq s} (\lambda_k-\lambda_j)
=
\prod_{1\leq j<k\leq s} \frac{t_k-t_j}{(1-t_k)(1-t_j)}
=
\frac{1}{\prod_{j=1}^s(1-t_j)^{s-1}}\prod_{1\leq j<k\leq s} (t_k-t_j),
\end{gather}
we obtain
\begin{gather*}
Z= \frac{(-1)^{s(N-s)}}{\prod_{k=1}^{s}(N-k)!}
\prod_{j=1}^s\frac{t_j^N}{(1-t_j)^{N-s+1}}
\prod_{1\leq j<k\leq s}\frac{1}{t_k-t_j}
\\
\hphantom{Z=}{} \times
\det\left[(-1)^{N-s+j-1}(N-s+j-1)!
\left(\frac{1-t_i}{t_i}\right)^{N-s+j}\big(1-t_i^{N-s+j}\big)\right]_{i,j=1,\dots,s}.
\end{gather*}
Simplifying factors coming from the determinant, we end up with
\begin{gather}\label{ZPP}
Z= \frac{(-1)^{\frac{s(s-1)}{2}}}{\prod_{1\leq j<k\leq s}(t_k-t_j)}
\det\big[
(1-t_i)^{j-1}\big(t_i^{s-j}-t_i^{N}\big)\big]_{i,j=1,\dots,s}.
\end{gather}
Note that this formula is in fact identical to~\eqref{Z_Kostov}
in the partial homogeneous limit, modulo factor $(1-t_i)^{s-1}$ for the entries
of the matrix, see also~\eqref{lateprods}.

In the fully homogeneous case, the similar calculation in the case of
\eqref{Zfullhom} yields
\begin{gather}\label{Zhomfin}
Z=\frac{1}{\prod_{k=1}^{s}(N-k)!\prod_{k=1}^{s-1}k!}
\det\big[(N-s+ i + j - 2 )!
\big(1 - t^{N-s + i + j - 1}\big)\big]_{i,j=1,\dots,s}.
\end{gather}
Note that, taking the limit $t_1,\dots,t_s\to t$ in~\eqref{ZPP} one obtains
the determinant of a different matrix; it can be easily shown, however, that it
can be reduced to that in~\eqref{Zhomfin} by a successive subtraction of its
rows.

Thus, we have shown that the representations due to Foda--Wheeler~\eqref{Z_Foda}
and due to Kostov~\eqref{Z_Kostov} lead to essentially the same formula already in
the partial homogeneous limit. For the homogeneous model it
is given by a Hankel determinant, see~\eqref{Zhomfin}.

\subsection{The limit of semi-infinite lattice}\label{Sec23}

The model with pDWBC is interesting in that it admits
the limit $N\to\infty$ under a certain restriction on the weights~\cite{PP-19}.
Namely, for the model with symmetric weights \eqref{symmetric}
one have to set
$a=1$ and $b<1$, that is apparently fulfilled in~\eqref{weights}.
The limit $N\to\infty$ means that the right
boundary goes to infinity, so that the lattice becomes a
semi-infinite lattice strip, with $s$ rows. The
boundary conditions on the right boundary become effectively vanishing,
that is guaranteed by fact that the $a$-weight is normalized to one
and that the $b$-weight, which is now the parameter~$t$, satisfies $0 \leq t<1$.

For rather general settings but limiting to the rational weights, the following
important property holds. For the
six-vertex model on $s\times N$ lattice with pDWBC and
with the weights satisfying condition \eqref{bc=a} at each lattice vertex,
the partition function in the limit $N\to\infty$
is equal to one,
\begin{gather}\label{Z=1}
\lim_{N\to\infty} Z=1.
\end{gather}
The property \eqref{Z=1} can be seen as a consequence of stochasticity
expressed by the relation~\eqref{bc=a}, see, e.g.,~\cite{GS-92}. Indeed, the
$s$ paths all entering at the left boundary must all exit at the top
boundary, with the total probability equal to~$1$, as far as the total
probability is conserved at each lattice vertex.

It is interesting to see how \eqref{Z=1} can approached from the determinant
representations considered above. As far as representation~\eqref{Z_Foda} involves
the determinant of $N\times N$ matrix, it can hardly be used for this purpose
directly. On the other hand, representation \eqref{Z_Kostov} almost immediately
leads to the result, since
\begin{gather*}
\prod_{k=1}^N b(\lambda_i,\nu_k)
\underset{N\to \infty}{\longrightarrow} 0,\qquad i=1,\dots,s.
\end{gather*}
Hence, the second term in the determinant in \eqref{Z_Kostov} vanishes, and
the determinant boils down to the Vandermonde determinant exactly cancelling the
pre-factor in~\eqref{Z_Kostov}, that yields~\eqref{Z=1}.

It is useful to mention how this result can be obtained from the
formula for the partition function of the homogeneous model, given
by~\eqref{Zhomfin}. It is convenient to consider the determinant of
a slightly more general matrix, with the entries
\begin{gather*}
A_{ij}=(\alpha+i+j-2)!\big(1-t^{\alpha+i+j-1}\big).
\end{gather*}
When $\alpha$ is large, they behave as
\begin{gather*}
A_{ij}=
(\alpha+i+j-2)!\big(1 + O\big(\rme^{-\alpha|\log t|}\big)\big).
\end{gather*}
For $s$ kept finite as $\alpha\to\infty$, we have
\begin{gather*}
\det [A_{ij} ]_{i,j=1,\dots,s}
=
\det[(\alpha+i+j-2)!]_{i,j=1,\dots,s}
\big(1 + O\big(\rme^{-\alpha|\log t|}\big)\big),
\end{gather*}
and so the determinant can be evaluated using the fact that
the leading term of $A_{ij}$ is nothing but
the moment of the orthogonality measure of the Laguerre polynomials
$\big\{L_n^{(\alpha)}(x)\big\}_{n=0}^\infty$,
\begin{gather*}
(\alpha+i+j-2)!=\int_{0}^\infty x^{i+j-2} \rme^{-x} x^{\alpha}\rmd x.
\end{gather*}
The standard technique of evaluation of the determinant of a Hankel
matrix yields
\begin{gather*}
\det [(\alpha+i+j-2)! ]_{i,j=1,\dots,s} =\prod_{j = 0}^{s-1} (\alpha + j)! j!.
\end{gather*}
Setting here $\alpha=N-s$, and substituting the result in \eqref{Zhomfin}, we obtain~\eqref{Z=1}.

\section{Boundary one-point function}\label{section3}

In this section we introduce and compute the boundary one-point function. We start
with exposing a connection of the model with off-shell Bethe states, give a
multiple integral formula for the one-point function, and present
a conjectural explicit expression. In the remaining part of this section
we give a proof of this result.

\subsection{Relation to off-shell Bethe states}\label{Sec31}

\begin{figure}\centering

\usetikzlibrary{decorations.pathreplacing}

\begin{tikzpicture}[scale=.5]
\draw [semithick] (0.2,1)--(9.8,1);
\draw [semithick] (0.2,2)--(9.8,2);
\draw [semithick] (0.2,3)--(9.8,3);
\draw [semithick] (0.2,4)--(9.8,4);
\draw [semithick] (1,0.2)--(1,4.8);
\draw [semithick] (2,0.2)--(2,4.8);
\draw [semithick] (3,0.2)--(3,4.8);
\draw [semithick] (4,0.2)--(4,4.8);
\draw [semithick] (5,0.2)--(5,4.8);
\draw [semithick] (6,0.2)--(6,4.8);
\draw [semithick] (7,0.2)--(7,4.8);
\draw [semithick] (8,0.2)--(8,4.8);
\draw [semithick] (9,0.2)--(9,4.8);
\draw [thick] [->] (.6,1)--(.5,1);
\draw [thick] [->] (.6,2)--(.5,2);
\draw [thick] [->] (.6,3)--(.5,3);
\draw [thick] [->] (.6,4)--(.5,4);
\draw [thick] [->] (4,4.5)--(4,4.4);
\draw [thick] [->] (9.5,1)--(9.6,1);
\draw [thick] [->] (9.5,2)--(9.6,2);
\draw [thick] [->] (9.5,3)--(9.6,3);
\draw [thick] [->] (9.5,4)--(9.6,4);
\draw [thick] [->] (1,.5)--(1,.6);
\draw [thick] [->] (2,.5)--(2,.6);
\draw [thick] [->] (3,.5)--(3,.6);
\draw [thick] [->] (4,.5)--(4,.6);
\draw [thick] [->] (5,.5)--(5,.6);
\draw [thick] [->] (6,.5)--(6,.6);
\draw [thick] [->] (7,.5)--(7,.6);
\draw [thick] [->] (8,.5)--(8,.6);
\draw [thick] [->] (9,.5)--(9,.6);
\draw [decorate,decoration={brace}]
(.8,5.1) -- (4.2,5.1) node [midway,yshift=9pt] {$m$};

\end{tikzpicture}

\caption{Definition of the boundary one-point function $G_\downarrow(m)$:
probability of observing configurations with the arrow on the top
boundary at the $m$th vertical line to be pointing down.}
\label{fig-GsNM}
\end{figure}
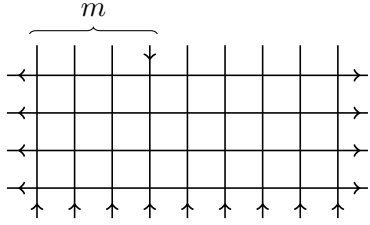

We define the boundary one-point function
as describing the probability of observing a down arrow,
at the $m$th, from the left, vertical line at the top boundary,
\begin{gather}\label{Gdowndef}
G_\downarrow(m)=\frac{1}{Z}
\sum_{\mathcal{C}\in\Omega^{\downarrow_m}}
\prod_{i=1,\dots, 6} w_i^{n_i(\mathcal{C})}.
\end{gather}
Here the summation is restricted to the
subset of configurations $\Omega^{\downarrow_m}\subset \Omega$
such that the arrow at the $m$th edge of the top boundary is fixed
to be pointed down, see Figure~\ref{fig-GsNM}.
In terms of lines, we thus require that this edge contains a solid line.
Since all configurations $\mathcal{C}$ contain exactly~$s$ solid lines,
the one-point function $G_\downarrow(m)=G_\downarrow(m;s,N)$ satisfies
the normalization condition (a~sum rule):
\begin{gather*}
\sum_{m=1}^{N}G_\downarrow(m)=s.
\end{gather*}

Thus, the boundary one-point function~\eqref{Gdowndef} describes
distribution of paths at the top boundary
where the free boundary conditions are imposed. Note that
in the special case $s=N$, where pDWBC become DWBC,
all edges on the top boundary contain solid lines, hence
$G_\downarrow(m;N,N)=1$. Our aim in this subsection is to show that
in the limit of the semi-infinite lattice, $N\to\infty$,
a nice explicit expression can be conjectured
for $G_\downarrow(m)$ as a function of the parameter~$t$ of
the homogeneous model Boltzmann weights~\eqref{weights}.

The key ingredient of our consideration here will be a connection of the
six-vertex model with pDWBC with off-shell Bethe states.
To outline this connection, and for a latter use, we need to give an operator formulation of the model in the framework of the QISM \cite{KBI-93}.

We start with introducing correspondence between arrow states
and the spin-up and spin-down vectors forming a basis of the
space $\mathbb{C}^2$ as follows:
\begin{gather}\label{arrows-spins}
\begin{split}
\uparrow, \rightarrow\quad \Longleftrightarrow\quad
\ket{\uparrow}\ \text{or}\ \bra{\uparrow},
\\
\downarrow, \leftarrow \quad \Longleftrightarrow\quad
\ket{\downarrow}\ \text{or}\
\bra{\downarrow}.
\end{split}
\end{gather}
The ket vectors are column vectors
\begin{gather*}
\kua =
\begin{pmatrix}
1 \\ 0
\end{pmatrix},
\qquad
\kda =
\begin{pmatrix}
0 \\ 1
\end{pmatrix},
\end{gather*}
and the bra vectors are the corresponding row vectors,
\begin{gather*}
\bra{\uparrow} =
\begin{pmatrix}
1 & 0
\end{pmatrix},
\qquad
\bra{\downarrow} =
\begin{pmatrix}
0 & 1
\end{pmatrix}.
\end{gather*}

Next, to the vertex being at the intersection of $k$th column and
$j$th row we associate an operator $L_{j k}(\lambda_j,\nu_k)$. Recall that
$\lambda_j$, $j=1,\dots,s$, and $\nu_k$, $k=1,\dots,N$, are the parameters
of the inhomogeneous version of the model, introduced in
Section~\ref{Sec22}. The $L$-operator
acts non-trivially in the direct product of two vector
spaces $\Cbb^2$: the ``horizontal'' space $\Hcal_j = \Cbb^2$
(associated with the $j$-th row) and the ``vertical'' space
$\Vcal_k = \Cbb^2$ (associated with the $k$th column).
Graphically, the $L$-operator acts from top to bottom and from right to
left. In other words, we define the $L$-operator as a matrix of the
Boltzmann weights of the vertex, with the arrow states on the top and right
edges as ``in'' indices, and those on the left and bottom edges, as ``out'' ones.

For the arbitrary weights $a_{jk}$, $b_{jk}$, and $c_{jk}$, the $L$-operator reads
\begin{gather*}
L_{jk}(\lambda_j, \nu_k)=
a_{jk}\frac{1+\tau_j^z\sigma_k^z}{2}+b_{jk}\frac{1-\tau_j^z\sigma_k^z}{2}+
c_{jk}\big(\tau_j^{+}\sigma_k^{-}+\tau_j^{-}\sigma_k^{+}\big),
\end{gather*}
where $\tau_j^{\pm,z}$ ($\sigma_k^{\pm,z}$) are the Pauli spin-$1/2$ operators
associated with the space $\Hcal_j$ ($\Vcal_k$). Specifically, for the
weights \eqref{1bc}, one has
\begin{gather*}
L_{j k} (\lambda_j, \nu_k) =
b(\lambda_j, \nu_k)
+ c(\lambda_j, \nu_k)\mathcal{P}_{jk},
\end{gather*}
where $\mathcal{P}_{jk}$ is the permutation operator, and
the functions $b(\lambda,\nu)$ and $c(\lambda,\nu)$ are defined in~\eqref{bc-lanu}.

Now consider a horizontal line, say $j$th, of our $s\times N$ square lattice, but
at the moment we let the states on the external horizontal edges take arbitrary values. The ``Boltzmann weight'' of the whole line can be given as
an ordered product of the $L$-operators, usually called
in the QISM the monodromy matrix:
\begin{align*}
T_j^\mathrm{V} (\lambda_j) = L_{j 1} (\lambda_j, \nu_1)
L_{j 2} (\lambda_j, \nu_2)
\cdots
L_{j N} (\lambda_j, \nu_N)
=
\begin{pmatrix}
A^\mathrm{V} (\lambda_j) & B^\mathrm{V} (\lambda_j)\\
C^\mathrm{V} (\lambda_j) & D^\mathrm{V} (\lambda_j)
\end{pmatrix}_{[\Hcal_j]}.
\end{align*}
Here, the $2\times 2$ matrix is written in the space $\Hcal_j$, as
indicated in the subscript. The operators $A^\mathrm
{V} (\lambda_j),\dots, D^\mathrm{V} (\lambda_j)$ depend on the
parameters $\nu_1,\dots,\nu_N$; we skip this dependence in the notation
for a simplicity. These operators act in the space
$\mathcal{V}=\otimes_{k=1}^N\Vcal_k$ and describe transition from a row to
row in the vertical direction, as indicated by the superscript.
Note that they contain information about which
horizontal line they correspond, namely the $j$th one,
only via their argument, $\lambda_j$.

Fixing the states on the external horizontal edges specifies which operator
out the four $A^\mathrm{V} (\lambda_j),\dots, D^\mathrm{V}
(\lambda_j)$ corresponds to the horizontal line. In the case of pDWBC, our
convention between arrow states and basis vectors \eqref{arrows-spins}
selects the $C^\textrm{V}$-operators for each line.
Denoting the basis vectors of the space
$\Vcal_k$ by $\ket{\uparrow_k}$ and $\ket{\downarrow_k}$, one can encode
the open boundary conditions on the top boundary by the state
\begin{equation*}
\ket{\Updownarrow^\mathrm{V}}
\equiv \otimes_{k=1}^N (\ket{\uparrow_k}+\ket{\downarrow_k}).
\end{equation*}
The boundary conditions on the bottom boundary correspond
to the state
\begin{equation*}
\bra{\Uparrow^\mathrm{V}}\equiv\otimes_{k=1}^N \bra{\uparrow_k}.
\end{equation*}
Thus, the partition function
of the six-vertex model with pDWBC can be written as the following
matrix element:
\begin{gather}\label{Zbracket}
Z=
\bra{\Uparrow^\mathrm{V}}C^\mathrm{V}(\lambda_s)\cdots C^\mathrm{V}(\lambda_1)\ket{\Updownarrow^\mathrm{V}}.
\end{gather}
The boundary one-point function \eqref{Gdowndef} can be defined as
\begin{gather*}
G_\downarrow(m)=\frac{1}{Z}
\bra{\Uparrow^\mathrm{V}}C^\mathrm{V}(\lambda_s)\cdots C^\mathrm{V}(\lambda_1)
\pi^\downarrow_m\ket{\Updownarrow^\mathrm{V}},
\end{gather*}
where $\pi^\downarrow_m$ denotes the projector to the spin-down state in the space
$\Vcal_m$, namely, $\pi^\downarrow_m=\frac{1}{2}(1-\sigma_m^z)$.

\begin{figure}\centering

\begin{tikzpicture}[scale=.5]
\draw [thick] (0.2,1)--(9.8,1);
\draw [thick] (0.2,2)--(9.8,2);
\draw [thick] (0.2,3)--(9.8,3);
\draw [thick] (0.2,4)--(9.8,4);
\draw [thick] (1,0.2)--(1,4.8);
\draw [thick] (2,0.2)--(2,4.8);
\draw [thick] (3,0.2)--(3,4.8);
\draw [thick] (4,0.2)--(4,4.8);
\draw [thick] (5,0.2)--(5,4.8);
\draw [thick] (6,0.2)--(6,4.8);
\draw [thick] (7,0.2)--(7,4.8);
\draw [thick] (8,0.2)--(8,4.8);
\draw [thick] (9,0.2)--(9,4.8);
\draw [thick] [->] (.5,1)--(.4,1);
\draw [thick] [->] (.5,2)--(.4,2);
\draw [thick] [->] (.5,3)--(.4,3);
\draw [thick] [->] (.5,4)--(.4,4);
\draw [thick] [<-] (1,4.5)--(1,4.4);
\draw [thick] [->] (2,4.5)--(2,4.4);
\draw [thick] [->] (3,4.5)--(3,4.4);
\draw [thick] [<-] (4,4.5)--(4,4.4);
\draw [thick] [->] (5,4.5)--(5,4.4);
\draw [thick] [<-] (6,4.5)--(6,4.4);
\draw [thick] [<-] (7,4.5)--(7,4.4);
\draw [thick] [<-] (8,4.5)--(8,4.4);
\draw [thick] [->] (9,4.5)--(9,4.4);
\draw [thick] [->] (9.5,1)--(9.6,1);
\draw [thick] [->] (9.5,2)--(9.6,2);
\draw [thick] [->] (9.5,3)--(9.6,3);
\draw [thick] [->] (9.5,4)--(9.6,4);
\draw [thick] [->] (1,.5)--(1,.6);
\draw [thick] [->] (2,.5)--(2,.6);
\draw [thick] [->] (3,.5)--(3,.6);
\draw [thick] [->] (4,.5)--(4,.6);
\draw [thick] [->] (5,.5)--(5,.6);
\draw [thick] [->] (6,.5)--(6,.6);
\draw [thick] [->] (7,.5)--(7,.6);
\draw [thick] [->] (8,.5)--(8,.6);
\draw [thick] [->] (9,.5)--(9,.6);
\node at (2,5.5) {$r_1$};
\node at (3,5.5) {$r_2$};
\node at (5,5.5) {$r_3$};
\node at (9,5.5) {$r_4$};

%
\end{tikzpicture}
\caption{Definition of the partition function $Z_{r_1,\dots,r_s}$:
the down arrows at the top boundary are fixed at the positions
$1\leq r_1<\dots<r_s\leq N$. Here, $N=9$, $s=4$, and $\{r_1,r_2,r_3,r_4\}=\{2,3,5,9\}$.}\label{fig-Ztop}
\end{figure}
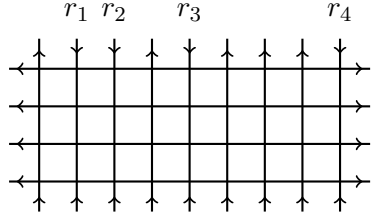

Now we are ready to establish the connection with the off-shell Bethe states.
For this end, we introduce one more object, namely the partition
function of the model on the $s\times N$ lattice with the down arrows located
at the positions $r_1,\dots,r_s$ on the top boundary, see Figure~\ref{fig-Ztop}.
Let us denote this partition function as $Z_{r_1,\dots,r_s}$. Similarly to
\eqref{Zbracket}, we can write
\begin{gather}\label{Zrs}
Z_{r_1,\dots,r_s}=
\bra{\Uparrow^\mathrm{V}}
C^\mathrm{V}(\lambda_s)\cdots C^\mathrm{V}(\lambda_1)\sigma_{r_1}^{-}\cdots\sigma_{r_s}^{-}
\ket{\Uparrow^\mathrm{V}}.
\end{gather}
To establish of a connection of the partition function $Z$ with $Z_{r_1,\dots,r_s}$, let us consider the vector~$\ket{\Updownarrow^\mathrm{V}}$ in~\eqref{Zbracket}.
Apparently, only the states with $s$ down spins contribute in~\eqref{Zbracket},
and therefore one can replace the vector $\ket{\Updownarrow^\mathrm{V}}$ by the sum over such states:
\begin{gather*}
\ket{\Updownarrow^\mathrm{V}}\Longrightarrow \sum_{1\leq r_1<\dots<r_s\leq N}^{} \sigma_{r_1}^{-}\cdots\sigma_{r_s}^{-} \ket{\Uparrow^\mathrm{V}}.
\end{gather*}
In other words, we have
\begin{gather*}
Z=\sum_{1\leq r_1<\dots<r_s\leq N}^{} Z_{r_1,\dots,r_s}.
\end{gather*}
The connection of $Z_{r_1,\dots,r_s}$ with the
boundary one-point function is slightly more subtle~\cite{CGP-21},
\begin{gather}\label{GsumZtop}
G_\downarrow(m)=Z^{-1}
\sum_{l=1}^{s}
\sum_{1\leq r_1<\dots<r_{l-1}< m< r_{l+1}<\dots<r_s\leq N}^{}
Z_{r_1,\dots,r_{l-1},m,r_{l+1},\dots,r_s},
\end{gather}
where the inner summation, over the values of $r$'s, is performed with~$m$ kept fixed.

Coming back to formula \eqref{Zrs}, the main observation one can make
about it is that by its form it hints at the fact that
$Z_{r_1,\dots,r_s}$ is nothing but a component of the off-shell Bethe
state in the sector with $s$ particles.
For a sake of simplicity, and also because we are
interested in the homogeneous model, we illustrate this fact for the partially
homogeneous model, which we define as in Section~\ref{Sec22}, by
setting $\nu_1,\dots,\nu_s=0$. Then introduce variables
\begin{gather*}
t_j=b(\lambda_j,0),\qquad j=1,\dots,s.
\end{gather*}
The correspondence between
the algebraic and coordinate versions of Bethe ansatz~\cite{IKR-87} (see also
\cite[Chapter~VII, Appendix~2]{KBI-93},),
implies that $Z_{r_1,\dots,r_s}$ in the coordinate form reads:
\begin{gather}\label{Boff}
Z_{r_1,\dots,r_s}
=\prod_{j=1}^{s}c_j\prod_{1\leq j<k\leq s}^{}\frac{1}{t_k-t_j}
\sum_{\sigma}(-1)^{[\sigma]} \prod_{j=1}^{s}
t_{\sigma_j}^{r_j-1} \prod_{1\leq j<k\leq s}^{} (1-2 t_{\sigma_j}+t_{\sigma_j}t_{\sigma_k}).
\end{gather}
Here $c_j\equiv(1-t_j)$, see \eqref{weights}. The
sum is performed over the permutations $\sigma\colon 1,\dots,s\mapsto \sigma_1,\dots,\sigma_s$,
and $[\sigma]$ denotes parity of $\sigma$.

Now, our task is to perform the homogeneous limit in the remaining set of
parameters, $t_1,\dots,t_s=t$, where $t$ defines the weights of the
homogeneous model, see \eqref{weights}. This not a quite straightforward
calculation, since there are singularities coming from the first factor
in \eqref{Boff} in the limit. The result can be formulated as follows.

\begin{Proposition}
The partition function $Z_{r_1,\dots,r_s}$ for the homogeneous
model can be represented as the $s$-fold contour integral
\begin{gather*}
Z_{r_1,\dots,r_s}=
\oint_{C_t}^{}\dots\oint_{C_t}^{}
\prod_{j=1}^{s}\frac{z_j^{r_j-1}}{(z_j-t)^s}
\prod_{1\leq j<k\leq s} p(z_j,z_k)\, \frac{\rmd z_1\cdots\rmd z_s}{(2\pi\rmi)^s},
\end{gather*}
where $C_t$ denotes a small simple counter-clockwise oriented contour around
the point $z=t$, and
\begin{gather*}
	p(x,y)=(x-y)(1-2x+x y).
\end{gather*}
\end{Proposition}

The proof can be found in \cite{CP-12}, see also \cite[Appendix~A]{CGP-21}.

The main practical usefulness of the representation in terms of
multiple contour integral is related to the fact that the residues
can be efficiently computed with the help of symbolic manipulation software.
The same also applicable to the sum in~\eqref{GsumZtop}, especially when
$N=\infty$. Recalling that $Z|_{N=\infty}=1$, see the discussion in Section~\ref{Sec23},
we have therefore been tempted to compute the boundary one-point function for the
case of semi-infinite lattice strip using~\eqref{GsumZtop} at $N=\infty$ for
some small values of $s=1,2,\dots$, in the hope that the results
will appear suitable for guessing of a general expression.

Indeed, performing this calculation, and denoting $g(m;s)\equiv G_\downarrow(m;s,\infty)$
with $n\equiv m-1=0,1,\dots$, we find for $s=1,\dots,5$, respectively:
\begin{gather*}
g(m,1)=(1-t)t^n,\\
g(m,2)=(1-t)t^{n-1}\big\{n(1-t)^2+(1+t)t\big\},\\
g(m,3)=(1-t) t^{n-2}
\left\{\frac{n(n-1)}{2}(1-t)^4+n(1-t)^2 t (1+2t)+t^2\left(1+t+t^2\right)\right\},\\
g(m,4) =
(1-t) t^{n-3} \bigg\{\frac{n(n-1)(n-2)}{6} (1-t)^6+\frac{(n-1)n}{2}(1-t)^4 t (1+3t)
\\ \hphantom{g(m,4) =}{}
+n (1-t)^2 t^2 \big(1+2t+3 t^2\big)
+t^3 \big(1+t+t^2+t^3\big)\bigg\},
\notag\\
g(m,5) =
(1-t) t^{n-4} \bigg\{\frac{n(n-1)(n-2)(n-3)}{24}(t-1)^8
 \\ \hphantom{g(m,5) =}{}
+\frac{n(n-1)(n-2)}{6}(1-t)^6 t (1+4t)
+\frac{n(n-1)}{2}(1-t)^4 t^2 \big(1+3t+6t^2\big)
 \\
\hphantom{g(m,5) =}{}
+n (1-t)^2t^3 \big(1+2t+3t^2+4t^3\big)
+t^4 \big(1+t+t^2+t^3+t^4\big)\bigg\}.
\end{gather*}
Inspecting these expressions, it is not difficult
to guess a general formula
\begin{gather*}
g(m,s)=(1-t)t^{n-s+1}
\sum_{j=0}^{s-1} \binom{n}{j} (1-t)^{2j}t^{s-j-1}\frac{1}{j!} \partial_t^{j}
\frac{1-t^s}{1-t}.
\end{gather*}
In this conjectured expression, however, we silently assume that~$s$ is small enough in
comparison with $n$. This restriction in fact can be relaxed, giving a formula, valid for
generic values of~$s$. Restoring the original notation in terms of~$m$, we thus have the
following.

\begin{Proposition}\label{Prop-main-result}
The boundary one-point function
of the homogeneous model
on the semi-infinite lattice, $g(m,s)=G_\downarrow(m;s,\infty)$, can be given in the form
\begin{gather}\label{Gconjecture}
g(m,s)= \sum_{j\geq 0}^{} \binom{m-1}{j} (1-t)^{2j+1}t^{m-1-j}\frac{1}{j!} \partial_t^{j}
\frac{1-t^s}{1-t}.
\end{gather}
\end{Proposition}
Note that the upper limit in the sum in \eqref{Gconjecture} is in fact
equal to $\min(m,s)-1$, since either
the binomial coefficient or the last factor terminates the sum.

Our aim in the remaining part of this section is to prove the assertion of
Proposition~\ref{Prop-main-result}.

\subsection{The QISM calculations}\label{Sec32}

To compute the boundary one-point function, we will use an
operator formulation of the model
in terms of ``horizontal'' monodromy matrix rather than the
``vertical'' one, used above.
Namely, we construct the monodromy matrix by considering a product
of the $L$-operators along the vertical line,
\begin{gather*}
T_k^\mathrm{H} (\nu_k) = L_{s k} (\lambda_s, \nu_k)
\cdots
L_{2 k} (\lambda_2, \nu_k)
L_{1 k} (\lambda_1, \nu_k)
=
\begin{pmatrix}
A^\mathrm{H} (\nu_k) & B^\mathrm{H} (\nu_k)\\
C^\mathrm{H} (\nu_k) & D^\mathrm{H} (\nu_k)
\end{pmatrix}_{[\Vcal_k]},
\end{gather*}
where the subscript recalls that this is a matrix in the space $\Vcal_k$.
The operators $A^\mathrm{H}\!(\nu_k),{\dots},D^\mathrm{H}\!(\nu_k)$ act in the space
$\Hcal=\otimes_{j=1}^s\Hcal_j$ and describe the transition from a column to
column in the horizontal direction, as indicated by the superscript.
Note that they contain information about which
vertical line they correspond, namely, the~$k$th one,
only via their argument,~$\nu_k$.

According to our conventions, see \eqref{arrows-spins}, the states
\begin{gather*}
\ket{\Ua^\mathrm{H}} = \otimes_{j = 1}^s \ket{\ua_j},
\qquad
\bra{\Da^\mathrm{H}} = \otimes_{j = 1}^s \bra{\da_j},
\end{gather*}
where $\ket {\ua_j}$ and $\bra{\da_j}$ are the basis
vectors in $\Hcal_j$, encode the boundary conditions
on the right and left boundaries, respectively.
The partition function of the six-vertex model on the $s \times N$ square
lattice with pDWBC can be defined as the matrix element:
\begin{gather*}
Z =
\bra{\Da^\mathrm{H}}
\prod_{k = 1}^N \big(A^\mathrm{H} (\nu_k) + B^\mathrm{H} (\nu_k)\big)
\ket{\Ua^\mathrm{H}}.
\end{gather*}
Here, the factors are ordered from the left to the right as~$k$ increases,
according to our convention on how the parameters $\nu_1,\dots,\nu_N$ are assigned to
the vertical lines of the lattice (see Section~\ref{Sec22}), though this
is not essential as we show below that all these operators commute with each other.

As for the one-point function $G_\downarrow (m)$,
we will use the fact that equivalently
we can consider the probability of having an up arrow at the given edge,
\begin{gather}\label{Gup1Gdn}
G_\uparrow(m)=1-G_\downarrow(m).
\end{gather}
Then,
\begin{gather}\label{G-bracket}
G_\uparrow (m) = Z^{-1}
\bra{\Da^\mathrm{H}}
\prod_{k=1}^{m-1}
\big(A^\mathrm{H} (\nu_k) + B^\mathrm{H} (\nu_k) \big)
A^\mathrm{H} (\nu_m)
\prod_{k = m+1}^{N} \big(A^\mathrm{H} (\nu_k) + B^\mathrm{H} (\nu_k)\big)\ket{\Ua^\mathrm{H}}.
\end{gather}
As we shall see, $G_\uparrow(m)$
can be computed in a rather straightforward manner;
we will commute operator $A^\mathrm{H} (\nu_m)$
to the left and act with it on the all-spins-down eigenstate,{\samepage
\begin{gather}\label{Avac}
\bra{\Da^\mathrm{H}} A^\mathrm{H} (\nu)
= \bra{\Da^\mathrm{H}} \prod_{j=1}^{s}b(\lambda_j,\nu).
\end{gather}
At the final stage we will turn back to $G_\downarrow(m)$.}

To compute the matrix element in~\eqref{G-bracket},
we need the commutation relations between the $A$- and $B$-operators.
They originate from the intertwining relation for the $L$-operators
\begin{gather}\label{eq:RLL}
R_{k k'} (\nu, \mu)
\big(L_{j k} (\lambda, \nu) \otimes L_{j {k'}} (\lambda, \mu)\big)
= \big(L_{j k'} (\lambda, \mu) \otimes	L_{j k} (\lambda, \nu)\big)
R_{k k'} (\nu, \mu).
\end{gather}
This relation is written as an operator equation in the direct product of spaces
$\Vcal_k \otimes \Vcal_{k'} \otimes \Hcal_j$.
The matrix $R_{k k'}(\nu,\mu)$ acts nontrivially in
$\Vcal_k \otimes \Vcal_{k'}$ and has the form
\begin{gather}\label{Rmat}
R_{k k'} (\nu, \mu) =
\begin{pmatrix}
f (\mu, \nu) & 0 & 0 & 0 \\
0 & 1 & g (\mu, \nu) & 0 \\
0 & g (\mu, \nu) & 1 & 0 \\
0 & 0 & 0 & f (\mu, \nu)
\end{pmatrix}_{[\Vcal_k \otimes \Vcal_{k'}]},
\end{gather}
where the functions $f (\mu, \nu)$ and $g (\mu, \nu)$ are (we follow the notation of \cite{KBI-93})
\begin{gather}\label{fg-funs}
f (\mu, \nu) = 1 + \frac{1}{\mu - \nu}, \qquad
g (\mu, \nu) = \frac{1}{\mu - \nu}.
\end{gather}
Relation \eqref{eq:RLL} implies a similar relation
for the monodromy matrix,
\begin{gather}\label{RTT}
R_{k k'} (\nu, \mu)
\big(T_k^\mathrm{H} (\nu) \otimes T_{k'}^\mathrm{H} (\mu) \big)
= \big(T_{k'}^\mathrm{H} (\mu) \otimes T_k^\mathrm{H} (\nu) \big) R_{k k'} (\nu, \mu).
\end{gather}

Relation \eqref{RTT} has an important property, known as
$gl_2$-invariance \cite{KS-82}, which can be seen as a consequence
of the particular form of the $R$-matrix \eqref{Rmat} and the identity
\begin{gather*}
f(\nu, \mu) - g(\nu, \mu) = 1.
\end{gather*}
It means that if one considers, instead of the monodromy matrix
$T_k^\mathrm{H}(\nu_k)$, the matrix
\begin{gather*}
\wt T_k^\mathrm{H}(\nu_k)=
K_\mathrm{L}T_k^\mathrm{H}(\nu_k) K_\mathrm{R},
\end{gather*}
where $K_\mathrm{L}$ and $K_\mathrm{R}$ are arbitrary $2\times 2$ matrices
with $c$-valued (non-operator) entries,
then relation~\eqref{RTT} remains intact.

From this place, we drop the superscript of operators and vectors,
and will write simply~$A(\nu)$ for~$A^\mathrm{H}(\nu)$, etc.
The sub-algebra spanned by the $A$- and $B$-operators contains the
commutativity properties
\begin{gather}\label{AABB}
\left[A (\nu), A (\mu)\right] =0,\qquad
\left[B (\nu), B (\mu)\right] =0,
\end{gather}
and the relations
\begin{gather}\label{ABBA}
\begin{split}
A(\nu) B(\mu) &= f(\nu,\mu) B(\mu) A(\nu) + g (\mu,\nu) B (\nu) A (\mu),
\\
B (\nu) A (\mu) &= f (\nu, \mu) A (\mu) B (\nu) + g (\mu, \nu) A (\nu) B (\mu).
\end{split}
\end{gather}
As it can be easily seen, the relations in \eqref{AABB} and \eqref{ABBA}
remain the same if we considers instead of the $A$- and $B$-operators any linear
combinations of them.
In particular, one can choose
\begin{gather}\label{wtAB}
\wt A(\nu)=A(\nu),\qquad
\wt B(\nu)=A (\nu) + B (\nu).
\end{gather}
This means, that \eqref{G-bracket} can be computed using the standard
methods of the QISM.

We will use the second relation in \eqref{ABBA} and move the
$A$-operator to left through a product of the $B$-operators.
In a completely standard
manner for QISM (see, e.g., \cite[Chapter~VII]{KBI-93}), we have
\begin{align}\label{BBBA}
\prod_{k = 1}^{m-1} B (\nu_k)
A(\nu_m)
&= \prod_{\substack{k = 1}}^{m-1}
f(\nu_k,\nu_m)\, A(\nu_m)\prod_{k = 1}^{m-1} B (\nu_k)
\notag\\ &\quad
+
\sum_{j = 1}^{m-1}
g (\nu_m, \nu_j)
\prod_{\substack{k = 1\\ k \neq j}}^{m-1}
f (\nu_k, \nu_j)\,
A(\nu_j)\prod_{\substack{k = 1\\ k \neq j}}^{m}
B (\nu_k)
\notag\\
&=\sum_{j=1}^m
\frac{g (\nu_m, \nu_j)}{f (\nu_m, \nu_j)}
\prod_{\substack{k = 1\\ k \neq j}}^{m}
f (\nu_k, \nu_j)\,
A(\nu_j)\prod_{\substack{k = 1\\ k \neq j}}^{m}
B (\nu_k),
\end{align}
where at the second step we have collected together
the ``wanted'' and ``unwanted'' terms into a~single sum,
using that $g(\nu,\nu)/f(\nu,\nu)=1$ \cite{BPZ-02}.

For the matrix element defining the boundary one-point function, see \eqref{G-bracket},
with \eqref{Avac}, \eqref{wtAB} and \eqref{BBBA} taken into account, one gets
\begin{gather*}
\bra{\Da}	\prod_{k = 1}^{m-1} \wt B(\nu_k)\, \wt A(\nu_m)
\prod_{k = m+1}^{N} \wt B(\nu_k)\ket{\Ua}\\
\qquad{}
=\sum_{j=1}^m
\prod_{l=1}^s b(\lambda_l,\nu_j)
\frac{g (\nu_m, \nu_j)}{f (\nu_m, \nu_j)}
\prod_{\substack{k = 1\\ k \neq j}}^{m}
f (\nu_k, \nu_j)\,
\bra{\Da}
\prod_{\substack{k = 1\\ k \neq j}}^{N}
\wt B (\nu_k) \ket{\Ua}.
\end{gather*}
Hence,
\begin{gather}\label{G=sumZ}
G_\uparrow (m) = Z^{-1}
\sum_{j=1}^m
\prod_{l=1}^s b(\lambda_l,\nu_j)
\frac{g (\nu_m, \nu_j)}{f (\nu_m, \nu_j)}
\prod_{\substack{k = 1\\ k \neq j}}^{m}
f (\nu_k, \nu_j)\,
Z(\setminus \nu_j),
\end{gather}
where $Z(\setminus \nu_j)$ denotes the partition function on an
$s\times (N-1)$ lattice, with the sets of parameters $\lambda_1,\dots,\lambda_s$
and $\nu_1,\dots,\nu_{j-1},\nu_{j+1},\dots,\nu_N$.

\looseness=-1 Expression \eqref{G=sumZ} is suitable for using an explicit expression for the
partition function in terms of a determinant to obtain a similar expression
for the boundary one-point function. For example,
due to the Foda--Wheeler formula~\eqref{Z_Foda}, the sum in~\eqref{G=sumZ}
can be seen as result of developing along a row (which has only the first~$m$ entries
not equal to zero) of some determinant of an $N \times N$ matrix.
As a result, this leads to the representation in terms of
a ratio of two determinants. We have discussed
this procedure and the subsequent homogeneous limit in details in~\cite{MP-19}.

We will consider here instead application of the formula by Kostov \eqref{Z_Kostov}.
This yields 	
\begin{gather*}
G_\uparrow (m) = Z^{-1}
\sum_{j=1}^m
\prod_{l=1}^s b(\lambda_l,\nu_j)
\frac{g (\nu_m, \nu_j)}{f (\nu_m, \nu_j)}
\prod_{\substack{k = 1\\ k \neq j}}^{m}
f (\nu_k, \nu_j)
\\
\hphantom{G_\uparrow (m) =}{} \times
\prod_{1\leq j<k\leq s}
\frac{1}{\lambda_k-\lambda_j}
\det\left[\lambda_i^{l-1}-(\lambda_i+1)^{l-1}
\prod_{\substack{k=1\\ k\ne j}}^N b(\lambda_i,\nu_k)\right]_{i,l=1,\dots,s}.
\end{gather*}
Let us investigate this expression in some detail. Using \eqref{fg-funs}, we have
\begin{align*}
\frac{g (\nu_m, \nu_j)}{f (\nu_m, \nu_j)}
\prod_{\substack{k = 1\\ k \neq j}}^{m}
f (\nu_k, \nu_j)
&=
\frac{1}{\nu_m-\nu_j+1}\prod_{\substack{k = 1\\ k \neq j}}^{m}
\frac{\nu_k-\nu_j+1}{\nu_k-\nu_j}
\notag\\
&=\frac{\prod_{k=1}^{m-1}(\nu_j-\nu_k-1)}{\prod_{\substack{k = 1\\ k \neq j}}^{m}
(\nu_j-\nu_k)},\qquad j=1,\dots,m,
\end{align*}
and so for a trial function $F(\nu)$, regular at the
points $\nu=\nu_1,\dots,\nu_m$, we can write
\begin{gather*}
\sum_{j=1}^m
\frac{\prod_{k=1}^{m-1}(\nu_j-\nu_k-1)}{\prod_{\substack{k = 1\\ k \neq j}}^{m}
(\nu_j-\nu_k)} F(\nu_j)=
\oint_{C_{\nu_1,\dots,\nu_m}}
\frac{\prod_{j=1}^{m-1}(\nu-\nu_j-1)}{\prod_{j=1}^{m}(\nu-\nu_j)}
F(\nu) \frac{\rmd \nu}{2\pi\rmi},
\end{gather*}
where $C_{\nu_1,{\dots},\nu_m}\!$ denotes a
simple counter-clockwise oriented contour enclosing
the points $\nu_1,{\dots},\nu_m\!$ and no other singularity of the integrand.
Hence,
\begin{gather*}
\begin{split}&
G_\uparrow (m) = Z^{-1}
\prod_{1\leq j<k\leq s}\frac{1}{\lambda_k-\lambda_j}
\oint_{C_{\nu_1,\dots,\nu_m}}
\frac{\prod_{j=1}^{m-1}(\nu-\nu_j-1)}{\prod_{j = 1}^{m}(\nu-\nu_j)}
\prod_{j=1}^s b(\lambda_j,\nu)
\\
& \hphantom{G_\uparrow (m) =}{} \times
\det\left[\lambda_i^{j-1}-(\lambda_i+1)^{j-1}
\frac{\prod_{\substack{k=1}}^N b(\lambda_i,\nu_k)}{b(\lambda_i,\nu)}
\right]_{i,j=1,\dots,s}
\frac{\rmd \nu}{2\pi\rmi}.
\end{split}
\end{gather*}

A nice property of the last expression is that the partial homogeneous
limit, namely, where $\nu_1,\dots,\nu_N\to 0$, can be readily taken,
with the result
\begin{gather}
G_\uparrow (m) = Z^{-1}
\prod_{1\leq j<k\leq s}\frac{1}{\lambda_k-\lambda_j}
\oint_{C_{0}}
\frac{(\nu-1)^{m-1}}{\nu^m}
\prod_{j=1}^s b(\lambda_j,\nu)\nonumber\\
\hphantom{G_\uparrow (m) =}{} \times
\det\left[\lambda_i^{l-1}-(\lambda_i+1)^{l-1}
\frac{[b(\lambda_i,0)]^N}{b(\lambda_i,\nu)}
\right]_{i,l=1,\dots,s}
\frac{\rmd \nu}{2\pi\rmi},\label{Gupparhom}
\end{gather}
where $C_0$ is a contour around the origin. We transform~\eqref{Gupparhom}, first,
by changing the integration variable $\nu\mapsto w$, by
\begin{gather*}
w=\frac{\nu-1}{\nu}.
\end{gather*}
The contour in the complex $w$-plane, is a contour surrounding the
point $w=\infty$ in the counter-clockwise direction, that is, it has
clockwise orientation around the origin. Second, we switch from the parameters
$\lambda_j$ to the parameters $t_j=b(\lambda_j,0)$, $j=1,\dots,s$. In particular,
we have
\begin{gather}\label{tautw}
b(\lambda_j,\nu)= \frac{1-2t_j+t_j w}{w-t_j}\equiv\tau(t_j,w).
\end{gather}
As a result, \eqref{Gupparhom} becomes
\begin{gather*}
G_\uparrow (m) = - Z^{-1}
\prod_{1\leq j<k\leq s}\frac{1}{t_k-t_j}
\oint_{C_{\infty}}
\frac{w^{m-1}}{1-w}
\prod_{j=1}^s \tau(t_j,w)
\\
\hphantom{G_\uparrow (m) =}{} \times
\det\left[(1-t_i)^{s-j}\left(t_i^{j-1}-\frac{t_i^N}{\tau(t_i,w)}\right)
\right]_{i,j=1,\dots,s}
\frac{\rmd w}{2\pi\rmi},
\end{gather*}
where $C_\infty$ denotes a large contour, counter-clockwise oriented around
the origin, and the overall minus sign is due to the change of the
orientation. We now shrink the contour, and pick up the
contribution coming from the simple pole at~$w=1$. Noting that,
since $\tau(t_j,1)=1$, the residue at this point exactly coincides with
the value of $Z$, we thus arrive back at $G_\downarrow(m)$, see~\eqref{Gup1Gdn},
with the result
\begin{gather}
G_\downarrow (m) = Z^{-1}
\prod_{1\leq j<k\leq s}\frac{1}{t_k-t_j}
\oint_{C_{t_1,\dots,t_s}}
\frac{w^{m-1}}{1-w}
\prod_{j=1}^s \tau(t_j,w)\nonumber\\
\hphantom{G_\downarrow (m) =}{}
\times \det\left[(1-t_i)^{s-j}\left(t_i^{j-1}-\frac{t_i^N}{\tau(t_i,w)}\right)
\right]_{i,j=1,\dots,s}
\frac{\rmd w}{2\pi\rmi}.\label{Gdownparhom}
\end{gather}
Here, $C_{t_1,\dots,t_s}$ is a contour enclosing the points $w=t_1,\dots,t_s$.

\subsection{The limit of semi-infinite lattice}

In the limit $N\to\infty$ the boundary one-point function simplifies
significantly, as it can be already seen at the stage of the expression
\eqref{G=sumZ} where $Z\to1$.
In the partially inhomogeneous case, from~\eqref{Gdownparhom} it follows
that the one-point function $g(m,s)\equiv G_\downarrow(m;s,\infty)$
is given by
\begin{gather*}
g (m,s) =
\oint_{C_{t_1,\dots,t_s}}
\frac{w^{m-1}}{1-w}
\prod_{j=1}^s \tau(t_j,w)
\frac{\rmd w}{2\pi\rmi},
\end{gather*}
where $\tau(t,w)$ is defined in \eqref{tautw}.

In the fully homogeneous case, we have simply
\begin{gather}\label{gint1}
g(m,s)=\oint_{C_t}
\frac{w^{m-1}}{(1-w)}\left(\frac{1-2t+tw}{w-t}\right)^s
\frac{\rmd w}{2\pi \rmi}.
\end{gather}
We are now ready to finalize the proof of Proposition~\ref{Prop-main-result},
namely, we will show that~\eqref{Gconjecture} and~\eqref{gint1} are equivalent.

We start with writing \eqref{Gconjecture} in the form
\begin{gather}\label{g=G-G}
g(m,s) = h(m,0)-h(m,s),
\end{gather}
where
\begin{gather*}
h(m,s)\equiv\sum_{j\geq 0} \binom{m-1}{j} (1-t)^{2j+1}t^{m-1-j}\frac{1}{j!} \partial_t^{j}
\frac{t^s}{1-t}.
\end{gather*}
For $h(m,0)$, we have
\begin{gather}\label{Gisone}
h(m,0)=\sum_{j\geq 0}^{} \binom{m-1}{j} (1-t)^{j}t^{m-1-j}=1.
\end{gather}
For $h(m,s)$, using the Cauchy formula, we obtain
\begin{align*}
h(m,s)&=(1-t)t^{m-1}\sum_{j\geq 0}^{} \binom{m-1}{j}
\left[\frac{(1-t)^{2}}{t}\right]^j
\oint_{C_t}
\frac{z^{s}}{(1-z)(z-t)^{j+1}}
\frac{\rmd z}{2\pi \rmi}
\\
&= (1-t)
\oint_{C_t}
\frac{z^{s}\left(1-2t+tz\right)^{m-1}}{(1-z)(z-t)^{m}}
\frac{\rmd z}{2\pi \rmi}.
\end{align*}
Making the change of the integration variable $z\mapsto w$, by
\begin{gather*}
w=\frac{1-2t+tz}{z-t},
\end{gather*}
so that $z=(1-2t+tw)/(w-t)$,
and noting that the integration contour in $w$-plane is around the
point $w=\infty$, we shrink the contour, and obtain
\begin{align*}
h(m,s)&=\oint_{C_\infty}
\left(\frac{1-2t+tw}{w-t}\right)^{s}\frac{w^{m-1}}{w-1}
\frac{\rmd w}{2\pi \rmi}
\\ &
=1-\oint_{C_t}
\left(\frac{1-2t+tw}{w-t}\right)^{s}\frac{w^{m-1}}{1-w}
\frac{\rmd w}{2\pi \rmi}.
\end{align*}
The last expression, in virtue of \eqref{g=G-G} and \eqref{Gisone}, yields~\eqref{gint1}.

As a last comment,
we mention here that~\eqref{gint1} is also
equivalent to the following
representation for the one-point function in terms of the Jacobi
polynomials, obtained in~\cite{MP-19}:
\begin{gather}\label{GviaJP}
g(m,s)=
t^{m-s}\sum_{j=0}^{s}\binom{s}{j}(-t)^j
P_{s-1}^{(m-s,j-s)}(1-2t).
\end{gather}
Indeed, let us employ the Rodrigues formula for the Jacobi polynomials, which implies
\begin{gather*}
\JP{n}{\alpha}{\beta}{1-2t} =
\frac{1}{n!} t^{-\alpha} (1-t)^{-\beta}
\frac{\rmd^n}{\rmd t^n}
t^{n+\alpha} (1-t)^{n+\beta}.
\end{gather*}
Using the Cauchy formula, one can write
\begin{gather}\label{JPn}
\JP{n}{\alpha}{\beta}{1-2t} =
t^{-\alpha} (1-t)^{-\beta}
\oint_{C_t} \frac{z^{n+\alpha}(1-z)^{n+\beta}}
{(z-t)^{n+1}} \frac{\rmd z}{2\pi\rmi}.
\end{gather}
From \eqref{JPn}, for \eqref{GviaJP} we obtain
\begin{align*}
g(m,s)&=(1-t)^s \sum_{j=0}^s \binom{s}{j} \left(-\frac{t}{1-t}\right)^j
\oint_{C_t} \frac{z^{m-1}(1-z)^{j-1}}
{(z-t)^{s}} \frac{\rmd z}{2\pi\rmi}
\\
&= (1-t)^s\oint_{C_t} \frac{z^{m-1}}
{(1-z)(z-t)^{s}} \left(1-\frac{t}{1-t}(1-z)\right)^s \frac{\rmd z}{2\pi\rmi},
\end{align*}
and so \eqref{gint1} follows.

\section{Scaling in the semi-infinite lattice case}\label{section4}

In this section we study boundary one-point function on
the semi-infinity lattice in the limit, where
both the number of horizontal lines and the position of the down arrow
tends to infinity, with their ratio kept fixed,
\begin{gather}\label{mslarge}
s, m \to \infty, \qquad \frac{m}{s}=:\mu, \qquad \mu\in[0,\infty).
\end{gather}
It can be regarded as a scaling limit where the mesh size of the lattice tends to zero,
so that the lattice is scaled to the semi-infinite strip of unit width, with
$\mu$ being the coordinate along the $x$-axis.

We study the one-point function
in the limit \eqref{mslarge} by two methods: the standard
saddle-point method applied to the integral representation and
using a method based on an ordinary differential equation (ODE).

\subsection{The saddle-point analysis}\label{Sec41}

To simply a bit the saddle-point analysis of the integral
formula for the one-point function,
we rewrite \eqref{gint1} making the change
of the integration variable $w\mapsto z$, by
\begin{gather*}
w=t+(1-t)z,
\end{gather*}
so that the new representation contains integration around the origin,
with singularities of the integrand at the points $z=0,1,\infty$,
\begin{gather}\label{gint2}
g(m,s)=\oint_{C_0}\frac{ (t+(1-t)z )^{m-1} (1-t+tz )^s}{(1-z)z^s}
\frac{\rmd z}{2\pi \rmi}.
\end{gather}

We rewrite further this representation as follows
\begin{gather*}
g(m,s)=
(1+\lambda)\oint_{C_0}
\frac{\exp\{sF(z)\}}{(1-z) (\lambda+z )}
\frac{\rmd z}{2\pi \rmi},
\end{gather*}
with
\begin{gather*}
F(z)=\mu \log (\lambda+z )+\log (1+\lambda z ) -\log z-(\mu+1)\log(1+\lambda),
\end{gather*}
where $\mu=m/s$
and we have used the notation $\lambda=t/(1-t)$, see~\eqref{tbla}.
The saddle-point equation $F'(z)=0$ is
\begin{gather}\label{speq}
\mu (1+\lambda z)z=\lambda+z.
\end{gather}
It has two solutions
\begin{gather}\label{zpm}
z_\pm=\frac{1-\mu\pm\sqrt{(1-\mu)^2+4\mu \lambda^2}}{2\mu\lambda}.
\end{gather}
Recalling that $\lambda,\mu\in[0,\infty)$, one can see from \eqref{zpm} that
$z_{\mp}\lessgtr 0$ for all (positive) values of~$\mu$ and~$\lambda$. Using
\eqref{speq}, it can be shown that
\begin{gather*}
F''(z)=\frac{\lambda \mu\big(\mu z^2+1\big)}{z(\lambda+z)^2}, \qquad z=z_\pm,
\end{gather*}
and, since $z_{\mp}\lessgtr 0$, we also have $F''(z_{\mp})\lessgtr 0$.
From this it follows that the steepest descent contour
for the point~$z_{-}$ (respectively, $z_{+}$)~goes along (perpendicular to) the real axis.
Recalling that the original contour of integration is around the origin,
we conclude that only the point~$z_{+}$ contributes into the
saddle-point method approximation; there is no contour satisfying
the minimax principle in the case of the point~$z_{-}$
(see, e.g.,~\cite{dB-58,F-77}).\footnote{This agrees also with the
rather general statement (see, e.g., \cite[Chapter~IV, Section~5]{F-77}) that
asymptotic behavior of the integrals of the form $\oint_{C_0} z^{-s}f(z)\,\rmd z$, where
$f(z)=\sum_{n\geq 0}a_n z^n$ with all $a_n>0$ and satisfying $f(1)=1$,
is governed by just one and only one saddle point
located at the positive half of the real axis, with
the steepest descent contour perpendicular to the real axis.}

Focusing at the point $z_{+}$ we note that $z_{+}<1$, for $\mu>1$,
and $z_{+}>1$, for $\mu<1$, so that
the pole at the point $z=1$ contributes in the latter case. This has a crucial
consequence on the form of the leading term of the asymptotics, namely it means that
it has a Heaviside step-function behavior with the jump occurring at the value $\mu=1$.
As a result, in the limit $s\to\infty$, $m=\mu s$, the saddle-point method yields
\begin{gather}\label{spa}
g(m,s)= \theta(1-\mu)+\sgn(\mu-1) \frac{\rme^{-s \Phi_1(\mu)-\Phi_0(\mu)}}{\sqrt{s}}\big(1+O\big(s^{-1}\big)\big),
\end{gather}
where $\theta(1-\mu)$ is the Heaviside step function,
and the functions $\Phi_{1,0}(\mu)=\Phi_{1,0}(\mu;t)$,
defined as
\begin{gather*}
\Phi_1(\mu)\equiv -F(z_{+}),\qquad \Phi_0(\mu)\equiv
-\log\frac{(1+\lambda)}{|1-z_{+}|(\lambda+z_{+})\sqrt{2\pi F''(z_{+})}},
\end{gather*}
are given explicitly by
\begin{gather}
\Phi_1(\mu)=-\mu\log\mu -(1+\mu)\log\left(\frac{1+\mu +\sqrt{(\mu-1)^2+4\mu \lambda^2}}{2\mu(1+\lambda)}\right)
\nonumber\\
\hphantom{\Phi_1(\mu)=}{} +(1-\mu)\log\left(\frac{1-\mu +\sqrt{(\mu-1)^2+4\mu \lambda^2}}{2\mu\lambda}\right)\label{Phi1}
\end{gather}
and
\begin{gather}
\Phi_0(\mu)=\log\left|1-\frac{1-\mu+\sqrt{(1-\mu)^2+4\mu \lambda^2}}{2\mu \lambda}\right|-
\log(1+\lambda)\nonumber\\
\hphantom{\Phi_0(\mu)=}{} +
\frac{1}{2}\log\left(2\pi\mu\sqrt{(1-\mu)^2+4\mu\lambda^2}\right),\label{Phi0}
\end{gather}
respectively, and we recall that here $\lambda\equiv t/(1-t)$.

Note that \eqref{spa} describes asymptotics of the one-point function
in the limit \eqref{mslarge} for $\mu$ less or larger than one; it does not provide
any information about how it scales when $\mu$ is close to one.

To study the behavior of the one-point function in the vicinity of the value $\mu=1$, we
introduce new parameter
\begin{gather*}
v= \frac{m-s}{\sqrt{s}}, \qquad v\in \mathbb{R},
\end{gather*}
or, in other words, we set $\mu=1+v/\sqrt{s}$. Correspondingly, we set
$z=1+ w/\sqrt{s}$, $w\in\rmi\mathbb{R}$, so that, for large $s$,
\begin{gather*}
F\big(1+w/\sqrt{s}\big)=\frac{\lambda w^2}{(1+\lambda)^2s}+\frac{vw}{(1+\lambda)s}
+O\big(s^{-3/2}\big).
\end{gather*}
The one-point function in the leading order \big(with the corrections
of $O\big(s^{-1/2}\big)$\big) evaluates as
\begin{align*}
g(m,s) &\approx
-\int_{-\epsilon -\rmi\infty}^{-\epsilon +\rmi\infty}
\frac{\exp\big\{\frac{\lambda w^2}{(1+\lambda)^2}+\frac{vw}{(1+\lambda)}\big\}}{w}
\frac{\rmd w}{2\pi \rmi}
\\
&= \frac{1}{2}-\frac{1}{2\pi\rmi} \fint_{-\infty}^\infty
\frac{\exp\big\{{-}\lambda w^2+\rmi vw\big\}}{w}
\rmd w
\\
&= \frac{1}{2}-\frac{1}{\pi} \int_{0}^\infty
\frac{\sin vw}{w} \rme^{-\lambda w^2}
\rmd w,
\end{align*}
where the last integral can be expressed in terms of the error function,
\begin{gather*}
\frac{2}{\pi}\int_{0}^\infty
\frac{\sin 2x w}{w} \rme^{- w^2}\rmd w=\erf(x).
\end{gather*}
Recall that
\begin{gather*}
\erf(x)\equiv \frac{2}{\sqrt{\pi}}\int_{0}^{x}\rme^{-w^2}\rmd w,
\qquad \erf(x)\in [-1,1],\quad x\in\mathbb{R}.
\end{gather*}
Thus, the one-point function near the step scales as follows:
\begin{gather}\label{erf}
\lim_{s\to\infty} g\big(s+[v\sqrt{s}],s\big)
= \frac{1}{2}-\frac{1}{2}\erf\left(\frac{v}{2\sqrt{\lambda}}\right)
=\frac{1}{2}\erfc\left(\frac{v}{2\sqrt{\lambda}}\right).
\end{gather}
The corrections to \eqref{erf} are of $O\big(s^{-1/2}\big)$.

Summarizing, we have just obtained that in the leading
order the one-point function behaves as a Heaviside step function; near the
jump it is described by the complementary error function. It is important to
note that such a singular behavior is governed
by a location of the steepest descent
contour within the saddle-point method in the case of the integral
representation~\eqref{gint2}.

Recall, that we have found two solutions of the saddle-point equation
but the position of the original contour suggests that
one of these two points must be ignored and the method must be
applied only to the remaining point, which should determine
the leading term of asymptotics.
However, there exists also a simple pole in the integrand
which may contribute as soon as the original contour is deformed to the steepest
descent one. The residue at this pole taken with the proper sign
dictated by the orientation of the original contour is exactly $1$, and so
the leading Heaviside step-function behavior follows; the sub-leading
contribution is exponentially small and comes from the saddle point.

Hence, the whole result concerning the asymptotic behavior of the one-point function
relies on a very subtle consideration in the complex plane, which constitutes
the topological part of the saddle-point method. It is therefore very desirable
to have an independent verification of the obtained result.
It turns out that this is indeed possible thanks to a remarkable property
of the one-point function to obey some finite-difference and differential equations.
One of them is an ODE whose particular form admits construction of the required
asymtoptics in a rather robust manner. In the remaining part of this section
we derive these equations and give an alternative derivation of the
asymptotics from the ODE.

\subsection{Finite-difference and differential equations}

Here our aim is to prove that the one-point function
satisfies simple finite-difference relations
with respect to the discrete parameters, and, most importantly, it also satisfies
certain second-order ODE in the variable $t$.

\begin{Proposition}
The one-point function $g(m,s)\equiv G_\downarrow(m;s,\infty)$
satisfies the finite-difference relations
\begin{gather}\label{Ds}
\Delta_s g=(1-t)t^{m+s-2}
\Ftwoone{-s+1}{-m+1}{1}{\left(\frac{1-t}{t}\right)^2}
\end{gather}
and
\begin{gather}\label{Dm}
\Delta_m g=-s(1-t)^2 t^{m+s-3}
\Ftwoone{-s+1}{-m+2}{2}{\left(\frac{1-t}{t}\right)^2},
\end{gather}
where, e.g., $\Delta_s g:=g(m,s)-g(m,s-1)$.
Furthermore, as a function of $t$ it
obeys the following second-order ordinary differential equation
\begin{gather}
\big[1+2(s-m)(1-t)\big](1-t)(1-2t) t^2y''
\nonumber\\
\qquad{} +2\big[1-6t+6t^2-2m(1-t)^2(1-5t)
+2s(1-t)\big(1-5t+5t^2\big)
\nonumber\\
\qquad{}+2\big(s^2-m^2\big)(1-t)^2t\big] ty'
\nonumber\\
\qquad{} - \big\{6(1-2t)t-m\big(1+ 13t-34t^2+16t^3\big)+s\big(1+7t-22t^2+16t^3\big)
\nonumber\\
\qquad{} +\big[3(s-m)^2-4\big(2s^2-sm-m^2\big)t+8\big(s^2-m^2\big)t^2\big] (1-t)
\nonumber\\
\qquad{} +2(s-m)^3 (1-t)^2\big\} y=0,\label{ODE}
\end{gather}
where $y=y(t):=\partial_t g(m,s)$.
\end{Proposition}
\begin{proof}
We will prove these equations by order.
The first relation,~\eqref{Ds}, in fact can easily be seen
already from~\eqref{Gconjecture}.
Indeed, since the dependence on $s$ is contained only in the last factor in~\eqref{Gconjecture}, one readily gets
\begin{align*}
\Delta_s g&=
\sum_{j\geq 0} \binom{m-1}{j} (1-t)^{2j+1}t^{m-1-j}\frac{1}{j!} \partial_t^{j}
t^{s-1}
\\
&= \sum_{j\geq 0} \binom{m-1}{j} \binom{s-1}{j} (1-t)^{2j+1}t^{m-2-2j}
\\
& = (1-t)t^{m-2}\Ftwoone{-s+1}{-m+1}{1}{\left(\frac{1-t}{t}\right)^2}.
\end{align*}

To prove \eqref{Dm} it is more convenient to resort to the integral representation
\eqref{gint1}, from which we get
\begin{align*}
\Delta_m g&=
-\oint_{C_t}w^{m-2} \left(\frac{1-2t+tw}{w-t}\right)^s \frac{\rmd w}{2\pi\rmi}
\\
&=-t^{m+s-1}\oint_{C_1}w^{m-2} \left(1+\frac{(1-t)^2}{t^2}\frac{1}{w-1}\right)^s \frac{\rmd w}{2\pi\rmi}
\\
&=-t^{m+s-1} \sum_{j=1}^{s} \binom{s}{j}
\left(\frac{1-t}{t}\right)^{2j}\frac{1}{(j-1)!}\partial_w^{j-1} w^{m-2}\big|_{w=1}
\\
& = - s(1-t)^2t^{m+s-3}\Ftwoone{-s+1}{-m+2}{2}{\left(\frac{1-t}{t}\right)^2}.
\end{align*}

To obtain the ODE \eqref{ODE}, we differentiate with respect to $t$, say,
the integral representation~\eqref{gint1}, and also compute similarly
the finite differences of this representation in~$s$ and~$m$.
Comparison of the expressions shows that the quantity $y\equiv \partial_t g(m,s)$
can be written down in terms of $\Delta_s g$ and $\Delta_m g$ as follows:
\begin{gather*}
y=-\frac{s}{1-t}\Delta_s g
-\frac{m-1}{1-t}\Delta_m g.
\end{gather*}
Using \eqref{Ds} and \eqref{Dm} we can obtain an expression for $y$ in terms of
${}_2F_1$-functions. Recalling that our aim is to derive a differential equation for $y$,
we conclude that it certainly will be an outcome of the known equation for
the hypergeometric function. Therefore, the two ${}_2F_1$-functions in~\eqref{Ds} and~\eqref{Dm} need to be expressed in terms of some ${}_2F_1$-function
and its derivative. Using the known relations (see, e.g., \cite[Section~2.8]{E-81})
we can write
\begin{gather*}
\Ftwoone{-s+1}{-m+2}{2}{z}\\
\qquad{} =\frac{1}{s}
\Ftwoone{-s+1}{-m+1}{1}{z}
+\frac{1-z}{s(m-1)}\FtwoonePrime{-s+1}{-m+1}{1}{z}.
\end{gather*}
Hence,
\begin{gather}
y=\{m-1-(m+s-1)t\}t^{m+s-3}
\Ftwoone{-s+1}{-m+1}{1}{\left(\frac{1-t}{t}\right)^2}
\nonumber\\
\hphantom{y=}{} -(1-2t)(1-t)t^{m+s-5}
\FtwoonePrime{-s+1}{-m+1}{1}{\left(\frac{1-t}{t}\right)^2}.\label{yFF}
\end{gather}
From this expression one can readily compute~$y'$. Furthermore, using
the fact that the function
\begin{gather*}
F=\Ftwoone{-s+1}{-m+1}{1}{z}
\end{gather*}
satisfies the equation
\begin{gather*}
z(1-z)F''+[1+(m+s-3)z]F'-(m-1)(s-1)F=0,
\end{gather*}
one can eliminate $F''$ in the expression for $y'$, thus obtaining $y'$
as some linear combination of $F$ and $F'$, similarly to \eqref{yFF} for~$y$.
Solving this system of two relations between~$y$,~$y'$ and~$F$,~$F'$ for the latter pair, and requiring that the resulting
expressions are indeed related by
differentiation, one can obtain some linear equation connecting~$y$,~$y'$, and~$y''$.
In this way, we arrive at~\eqref{ODE}.
\end{proof}

\subsection{Construction of asymptotics from the ODE}\label{Sec43}

The asymptotics
of the one-point function can also be obtained
from the equation~\eqref{ODE} as a~solution identified by its $t\to 0$
expansion. We follow here the method originally proposed in~\cite{KP-16}.

We start with $t\to 0$ expansions, which can be easily obtained from
our main explicit formula for the one-point function $g=g(m,s)$,
see \eqref{Gconjecture}.
Assuming that $s<m$, we obtain
\begin{gather}\label{gslessm}
g=\binom{m-1}{s-1} t^{m-s}
\left\{1-\left(2s-\frac{m}{m-s+1}\right)t+O\left(t^2\right)\right\},
\end{gather}
and when $s>m$, we get
\begin{gather}\label{gmlesss}
g=1-\binom{s}{m-1} t^{s-m+1}
\left\{1
-\left(2m-2-\frac{m-1}{s-m+2}\right)t
+O\big(t^2\big)\right\}.
\end{gather}

Let us study how expressions \eqref{gslessm} and \eqref{gmlesss}
behave in the limit \eqref{mslarge}.
Using the $z\to\infty$ asymptotics of
the logarithm of the Gamma-function (see, e.g., \cite{E-81}),
\begin{gather*}
\log\Gamma(z+a)=
\left(z+a-\frac{1}{2}\right)\log z -z
+\frac{1}{2}\log 2\pi
\\
\hphantom{\log\Gamma(z+a)=}{}+\sum_{n=1}^\ell
\frac{(-1)^{n+1}B_{n+1}(a)}{n(n+1)z^n}+O\big(z^{-\ell-1}\big),
\qquad |\arg z|<\pi,
\end{gather*}
where $B_n(a)$ are the Bernoulli polynomials,
$B_n(a)=\sum_{k=0}^n \binom{n}{k}B_k a^{n-k}$
and $B_k$ are the Bernoulli numbers,
the asymptotics of the binomial coefficients
in~\eqref{gslessm} and~\eqref{gmlesss} can be readily computed up to any given order.
Here, we limit ourselves by the usual approximation of the Stirling formula, though
a more detailed analysis is possible along the lines explained below.

We first write the logarithm of \eqref{gslessm},
\begin{gather*}
\log g= (m-s)\log t +\log \binom{m-1}{s-1}
-\left(2s-\frac{m}{m-s+1}\right)t+O\big(t^2\big).
\end{gather*}
In the limit \eqref{mslarge} it can be written in the form
\begin{gather}\label{logg-mla}
\log g = \psi_1 s -\frac{1}{2}\log s +\psi_0+ O\big(s^{-1}\big),\qquad \mu>1,
\end{gather}
where
\begin{gather}\label{psi10}
\begin{split}
\psi_1&=(\mu-1)\log t+\mu\log\mu-(\mu-1)\log(\mu-1)-2t +O\big(t^2\big),
\\
\psi_0&=-\frac{1}{2}\log\big( 2\pi \mu(\mu-1)\big)+\frac{\mu}{\mu-1}t
+O\big(t^2\big).
\end{split}
\end{gather}
Essentially similarly, from \eqref{gmlesss} we have
\begin{gather*}
\log(1-g)=(s-m+1)\log t
+\log \binom{s}{m-1} -\left(2m-2-\frac{m-1}{s-m+2}\right)t
+O\big(t^2\big),
\end{gather*}
and so in the limit \eqref{mslarge},
\begin{gather}\label{logg-msm}
\log(1- g) = \wt\psi_1 s -\frac{1}{2}\log s +\wt\psi_0+ O\big(s^{-1}\big),\qquad \mu<1,
\end{gather}
where
\begin{gather}\label{wtpsi10}
\begin{split}
\wt\psi_1&=(1-\mu)\log t-\mu\log\mu-(1-\mu)\log(1-\mu)-2\mu t +O\big(t^2\big),
\\
\wt\psi_0&=\log t-\frac{1}{2}\log \left(2\pi \frac{(1-\mu)^3}{\mu}\right)
+\frac{2-\mu}{1-\mu}t
+O\big(t^2\big).
\end{split}
\end{gather}
Our aim below is to construct functions $\psi_1$, $\psi_0$ and
$\wt\psi_1$, $\wt\psi_0$ which possess $t\to 0$ asymptotics given by
\eqref{psi10} and \eqref{wtpsi10}, respectively.
In \eqref{logg-mla} and \eqref{logg-msm}
we have anticipated the fact that the term $-\frac{1}{2}\log s$
is exact, as it will be clearly seen below.

To study $\mu>1$ case, let us introduce an auxiliary function
\begin{gather*}
\sigma\equiv \partial_t \log g.
\end{gather*}
From \eqref{logg-mla}, the function $\sigma$ is of $O(s)$ in the limit \eqref{mslarge} at
$\mu>1$. Moreover, since
\begin{gather}\label{s1s2s3}
y=\sigma g, \qquad
y'=\big(\sigma^2+\sigma'\big)g,\qquad
y''= \big(\sigma^3+3\sigma\sigma'+\sigma''\big)g,
\end{gather}
from \eqref{ODE} it follows that the function $\sigma$, up to exponentially small
corrections, possesses an expansion in inverse powers of $s$, starting from
the first order term,
\begin{gather}\label{anzats}
\sigma = \sigma_1 s +\sigma_0 + \sigma_{-1} s^{-1}+\cdots.
\end{gather}
In particular, from \eqref{psi10} we see that, as $t\to 0$,
\begin{gather}\label{sigma1at0}
\sigma_1= \psi_1'=\frac{\mu-1}{t} -2 +O(t).
\end{gather}
Substituting \eqref{s1s2s3} and \eqref{anzats} into the differential equation \eqref{ODE},
and picking up the highest term in $s$, we
find that $\sigma_1$ must satisfy the following equation:
\begin{gather}\label{eqsigma1}
t^2(1-2t) \sigma_1^3 +2 t^2(1+\mu) \sigma_1^2 -(\mu-1)^2\sigma_1=0.
\end{gather}
The root, matching \eqref{sigma1at0}, is
\begin{gather}\label{sigma1}
\sigma_1= \frac{-t(1+\mu)+\sqrt{(\mu-1)^2(1-t)^2+4\mu t^2}}{t(1-2t)},\qquad \mu>1.
\end{gather}
Integrating this expression with respect to $t$
and fixing the integration constant to satisfy \eqref{psi10},
we find
\begin{gather*}
\psi_1= \int\sigma_1\rmd t=-\Phi_1(\mu)
\end{gather*}
where the quantity $\Phi_1(\mu)=\Phi_1(\mu;t)$ is given by~\eqref{Phi1}.
As far as the leading term of the expansion~\eqref{anzats} is established,
other terms can be constructed recursively up to any given order.
For example, for the term~$\sigma_0$
from~\eqref{ODE} we obtain
\begin{gather}
\sigma_0=
\frac{1}{(1-\mu )^2-4 (\mu +1) t^2 \sigma_1-3 (1-2 t) t^2 \sigma_1^2}
\bigg\{2 \left(1+\mu\right) t^2 \sigma_1'
\nonumber\\
\hphantom{\sigma_0=}{} +3 (1-2 t) t^2 \sigma_1 \sigma_1'
-\frac{3 (1-\mu )-4 (\mu +2)t+8 (\mu +1) t^2}{2 (1-t)}\sigma_1
\nonumber\\
\hphantom{\sigma_0=}{} +2\frac{(1-\mu )\big(1+5t^2\big)-(5-6 \mu ) t}{(1-\mu ) (1-t)}t \sigma_1^2
+\frac{(1-2 t)}{2 (1-\mu ) (1-t)} t^2 \sigma_1^3
\bigg\}.\label{eqsigma0}
\end{gather}
Substituting here \eqref{sigma1}, integrating, and fixing integration constant
to match the $t\to 0$ behavior in according to~\eqref{psi10}, we recover that
$\psi_0=\int \sigma_0\rmd t=-\Phi_0(\mu)$, where the function $\Phi_0(\mu)=\Phi_0(\mu;t)$
is given by~\eqref{Phi0}.

Coming back to the equation describing the leading term of the expansion~\eqref{anzats}, na\-me\-ly~\eqref{eqsigma1}, one can notice that the trivial root
$\sigma_1=0$ in fact describes this function for the values $\mu<1$.
(Recall that we assume $\mu$ to be positive, and
the third root, with the opposite sign of the square root expression in~\eqref{sigma1},
corresponds to the unphysical region where~$\mu$ is negative.)
Correspondingly, from~\eqref{eqsigma0} one can see that then $\sigma_0=0$;
moreover, all terms in the expansion~\eqref{anzats} vanish identically, so that
the function~$\sigma$ is given only by exponentially small terms. This is in agreement
with the fact that instead at $\mu<1$ we have the expansion~\eqref{logg-msm}.

To construct this expansion, we proceed essentially similarly as we did above
for $\mu>1$, namely, we introduce an auxiliary function
\begin{gather*}
\tilde \sigma \equiv \partial_t\log (1-g),
\end{gather*}
so that
\begin{gather}\label{s1s2s3tilde}
y=-\tilde\sigma (1-g), \qquad
y'=-\big(\tilde\sigma^2+\tilde\sigma'\big)(1-g),\qquad
y''=-\big(\tilde\sigma^3+3\tilde\sigma\tilde\sigma'+\tilde\sigma''\big)(1-g).
\end{gather}
From \eqref{ODE} it follows that $\tilde\sigma$, up to exponentially small
corrections, possesses an expansion of the form
\begin{gather}\label{anzatstilde}
\tilde\sigma = \tilde\sigma_1 s +\tilde\sigma_0 + \tilde\sigma_{-1} s^{-1}+\cdots.
\end{gather}
Substitution of \eqref{s1s2s3tilde} and \eqref{anzatstilde} into \eqref{ODE}
shows that the function $\tilde\sigma_1$ satisfies the same equation as
$\sigma_1$ does, namely, \eqref{eqsigma1}, and moreover, since from
\eqref{wtpsi10} is follows that, as $t\to0$,
\begin{gather*}
\tilde \sigma_1=\tilde\psi_1'=\frac{1-\mu}{t}-2\mu t+ O\big(t^2\big),
\end{gather*}
we conclude that $\tilde \sigma_1$ is given by \eqref{sigma1}, where instead we have to
assume that $\mu<1$. Similarly, the function $\tilde\sigma_0$
is given in terms of $\tilde\sigma_1$ by the expression \eqref{eqsigma0} in which
the replacement $\sigma_{1,0}\mapsto \tilde\sigma_{1,0}$ is to be made.
After substitution of the expression for $\tilde\sigma_1$, integrating, and fixing the
integration constant to match the $t\to 0$ behavior, see \eqref{wtpsi10},
one can find that $\tilde\psi_0=\int \tilde\sigma_0\rmd t=-\Phi_0(\mu)$,
where the function $\Phi_0(\mu)=\Phi_0(\mu;t)$ is given by \eqref{Phi0}.

As a result of this calculation, we thus obtain that
in the limit $s\to\infty$, $m=\mu s$, the one-point function is given by \eqref{spa},
in agreement with the saddle-point analysis.

Let us now consider the scaling properties of the one-point function
near the value $\mu=1$. We set $m=s+v\sqrt{s}$ and consider the large $s$ limit in
ODE~\eqref{ODE}. From this equation it follows that there exists a solution
of the form
\begin{gather*}
y=y_0+\frac{1}{\sqrt{s}}y_{-1}+\cdots,
\end{gather*}
where the leading term (corresponding to the coefficient of $s^{3/2}$ in \eqref{ODE})
must solve the equation
\begin{gather*}
y_0'-\frac{1}{2}\left\{\frac{1}{1-t}-\frac{3}{t}+\frac{v^2}{2t^2}\right\}y_0=0.
\end{gather*}
Therefore for the one-point function, $g'\equiv y$, in the leading order we obtain
\begin{gather}
g=C_1 \int_{t}^{1} \rme^{-\frac{v^2}{4w}} \frac{1}{w^{3/2}\sqrt{1-w}}\rmd w +C_2
= C_1\frac{2\sqrt{\pi}}{v} \rme^{-\frac{v^2}{4}} \erf\left(\frac{v}{2}\sqrt{\frac{1-t}{t}}\right) +C_2,\label{gatmu1ODE}
\end{gather}
where the integration constants may depends on $v$, $C_{1,2}=C_{1,2}(v)$.
They can be fixed by requiring that
\begin{gather*}
\lim_{t\to 0} g =
\begin{cases}
0, & v>0, \\
1, & v<0
\end{cases}
\end{gather*}
to match the leading $t\to 0$ asymptotic behavior of the one-point function
described by \eqref{gslessm} and \eqref{gmlesss}.
This gives $C_2=1/2$ and $C_1=-v \rme^{v^2/4}/4$, and, as a result, \eqref{gatmu1ODE}
reproduces~\eqref{erf}.

\section{Conclusion}\label{section5}

In this paper, we have computed the boundary one-point function of the
six-vertex model with partial domain wall boundary conditions in the case
of the rational Boltzmann weights. For the $s\times N$ lattice the result is
expressed in a determinantal form, see~\eqref{Gdownparhom}. In the limit of
the semi-infinite lattice, $N\to \infty$, it boils down to a simple explicit
expression, which can be given as the terminating series~\eqref{Gconjecture}
or as the contour integral~\eqref{gint1}. Furthermore, it turns out that
the one-point function can be determined as a polynomial solution
of certain second-order ODE, see~\eqref{ODE}.

Even though all these various representations for the one-point function can
be regarded as main results of the present paper, we would like to draw an
attention to the QISM calculations presented in Section~\ref{Sec32}. There, we
have shown that the one-point function can be directly obtained in terms of
a contour integral, provided that for the partition function of the
inhomogeneous model the Kostov determinant formula \eqref{Z_Kostov} is used.
This formula has a simple homogeneous limit in one of the two sets of the
parameters and it also has a simple large $N$ limit, which can also be done
before the homogeneous one. As a result, the derivation of the one-point
function for the model on the semi-infinite lattice presented here turns out
to be considerably simpler in comparison with that given in \cite{MP-19}.

Another point to which we would like to draw attention, is the existence of the ODE
which drives the one-point function. Even though this equation
is given by rather cumbersome expression \eqref{ODE}, it must be stressed that
it has appeared here in the context of a non-free-fermionic model.
Typically, examples of models where differential equations, usually non-linear ones,
describe correlation function are limited by those related to
dynamics of free fermions (see, e.g.,~\cite{KBI-93}).

The ODE which we have obtained for the one-point function is
with respect to the rapidity variable
of the weights, with the coefficients depending on the geometric parameters
of the one-point function in the semi-infinite lattice case, $m$ and $s$. Such an
equation is very interesting because it can be used to construct an
asymptotics in the limit of large $m$ and $s$ with the ratio $m/s= \mu$
fixed. We have showed how this can be done in Section~\ref{Sec43} using the
method developed in \cite{KP-16} on the example of the $\sigma$-form of the
sixth Painlev\'e equation.

An advantage of this method is that it allows for constructing an asymptotic expansion
recursively, starting from the leading term. Calculations are performed
in an almost purely algebraic manner.
Indeed, in the case of the contour integral \eqref{gint2},
the saddle-point method turns out to be very subtle in its topological part
(see Section~\ref{Sec41}): among the two saddle points one of
them appears to be completely irrelevant and the
leading term is determined by a~relative position of the second saddle point and
the simple pole of the integrand. In the case of the method based on the ODE \eqref{ODE},
the leading term immediately follows as a proper
root of the algebraic cubic equation \eqref{eqsigma1}.

\looseness=-1 One more point which definitely need to be discussed concerns the obtained
asymptotic behavior of the boundary one-point function
on the semi-infinite lattice in the scaling limit, $m,s\to\infty$ with $\mu=m/s$ kept fixed,
in which the mesh size of the lattice tends to zero.
In the leading order it is given
by the Heaviside step function, see~\eqref{spa}. In relation to this result
one may wonder about it interpretation in the context of limit shape phenomena and
what to expect if instead of the semi-infinite lattice the original
$s\times N$ lattice will be taken with the ratio $s/N$ kept fixed.

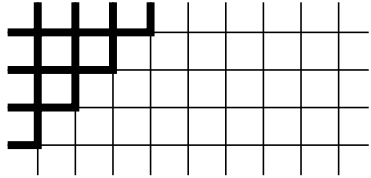
\begin{figure}\centering

\usetikzlibrary{decorations.pathreplacing}

\begin{tikzpicture}[scale=.5]
\draw [semithick] (0.2,1)--(9.8,1);
\draw [semithick] (0.2,2)--(9.8,2);
\draw [semithick] (0.2,3)--(9.8,3);
\draw [semithick] (0.2,4)--(9.8,4);
\draw [semithick] (1,0.2)--(1,4.8);
\draw [semithick] (2,0.2)--(2,4.8);
\draw [semithick] (3,0.2)--(3,4.8);
\draw [semithick] (4,0.2)--(4,4.8);
\draw [semithick] (5,0.2)--(5,4.8);
\draw [semithick] (6,0.2)--(6,4.8);
\draw [semithick] (7,0.2)--(7,4.8);
\draw [semithick] (8,0.2)--(8,4.8);
\draw [semithick] (9,0.2)--(9,4.8);
\draw [line width=3] (0.2,4)--(1,4)--(1,4.8);
\draw [line width=3] (0.2,3)--(1,3)--(1,4)--(2,4)--(2,4.8);
\draw [line width=3] (0.2,2)--(1,2)--(1,3)--(2,3)--(2,4)--(3,4)--(3,4.8);
\draw [line width=3] (0.2,1)--(1,1)--(1,2)--(2,2)--(2,3)--(3,3)--(3,4)--(4,4)--(4,4.8);
\end{tikzpicture}
\caption{The configuration without the $b$-weight vertices.}
\label{fig-limitshape}
\end{figure}

Using the description of the states in terms of the solid lines
(see Section~\ref{Sec21}), it is clear that the step-function behavior of the boundary
one-point function implies the dominance of the configurations such that
the first $s$ boundary edges on the vertical lines
are all occupied by the solid lines. Furthermore, there exists single configuration
characterized by absence of the $b$-weight vertices,
see Figure~\ref{fig-limitshape}, which,
for large $s$, together with very similar configurations
(containing small enough amount of $b$-weight vertices)
dominates over all other configurations.
In the $s\to\infty$ limit, this configuration describes
the limit shape. Thus, the step-function behavior of the boundary
one-point function can be regarded as a manifestation, even if rather trivial, but
the limit shape phenomenon occurring in the scaling limit.

It is also clear from the picture of Figure~\ref{fig-limitshape} that the
location of the right boundary has no effect on the limit shape. In other
words, it means that for the boundary one-point function the step-function
behavior should also be expected for the $s\times N$ lattice; however, the
corrections to this leading scaling behavior may depend on the given value of
$s/N$. This is also in agreement with the result of paper \cite{BL-15}
(see Theorem~1.3 and the discussion in Section~1.3 therein) that the free energy
of the model on the $s \times N$
lattice comes only from the `ground state' configuration,
which is exactly that shown in Figure~\ref{fig-limitshape}.

The irrelevance of the position of the right boundary for the
limit shape can be explained by peculiarity of the six-vertex model with the
rational weights. They belong to the ferroelectric regime on the phase
diagram, and therefore they are compatible with the ferroelectric order
induced by the boundary conditions in the case of pDWBC coming from the left,
bottom, and right boundaries.

To obtain nontrivial limit shape phenomena, one should take the model with
weights corresponding to the disordered or anti-ferroelectric regimes,
similarly to the DWBC case. However, these weights are trigonometric,
for which even finding a closed expression for the partition function
in the case of pDWBC is still an open problem in general. Definitely,
this deserves further study and we believe that many interesting results will be obtained
in this direction.

\subsection*{Acknowledgments}

The authors thank N.M.~Bogoliubov, F.~Colomo, N.~Reshetikhin, E.~Sobko for
stimulating discussions and the anonymous referees for valuable remarks.
This work was supported in part by the Russian Science Foundation,
grant \#~18-11-00297.

\pdfbookmark[1]{References}{ref}
\LastPageEnding


\begin{thebibliography}{99}
\footnotesize\itemsep=0pt

\bibitem{AR-05}
Allison D., Reshetikhin N., Numerical study of the 6-vertex model with domain
 wall boundary conditions, \href{https://doi.org/10.5802/aif.2144}{\textit{Ann. Inst. Fourier (Grenoble)}} \textbf{55}
 (2005), 1847--1869, \href{https://arxiv.org/abs/cond-mat/0502314}{arXiv:cond-mat/0502314}.

\bibitem{B-82}
Baxter R.J., Exactly solved models in statistical mechanics, Academic Press,
 Inc., London, 1982.

\bibitem{BR-20}
Belov P., Reshetikhin N., The two-point correlation function in the six-vertex
 model, \href{https://arxiv.org/abs/2012.05182}{arXiv:2012.05182}.

\bibitem{BL-14}
Bleher P., Liechty K., Random matrices and the six-vertex model, \textit{CRM
 Monograph Series}, Vol.~32, \href{https://doi.org/10.1090/crmm/032}{Amer. Math. Soc.}, Providence, RI, 2014.

\bibitem{BL-15}
Bleher P., Liechty K., Six-vertex model with partial domain wall boundary
 conditions: ferroelectric phase, \href{https://doi.org/10.1063/1.4908227}{\textit{J.~Math. Phys.}} \textbf{56} (2015),
 023302, 28~pages, \href{https://arxiv.org/abs/1407.8483}{arXiv:1407.8483}.

\bibitem{BPZ-02}
Bogoliubov N.M., Pronko A.G., Zvonarev M.B., Boundary correlation functions of
 the six-vertex model, \href{https://doi.org/10.1088/0305-4470/35/27/301}{\textit{J.~Phys.~A: Math. Gen.}} \textbf{35} (2002),
 5525--5541, \href{https://arxiv.org/abs/math-ph/0203025}{arXiv:math-ph/0203025}.

\bibitem{CGP-21}
Colomo F., Di~Giulio G., Pronko A.G., Six-vertex model on a finite lattice:
 {I}ntegral representations for nonlocal correlation functions,
 \href{https://doi.org/10.1016/j.nuclphysb.2021.115535}{\textit{Nuclear Phys.~B}} \textbf{972} (2021), 115535, 42~pages,
 \href{https://arxiv.org/abs/2107.13358}{arXiv:2107.13358}.

\bibitem{CP-09}
Colomo F., Pronko A.G., The arctic curve of the domain-wall six-vertex model,
 \href{https://doi.org/10.1007/s10955-009-9902-2}{\textit{J.~Stat. Phys.}} \textbf{138} (2010), 662--700, \href{https://arxiv.org/abs/0907.1264}{arXiv:0907.1264}.

\bibitem{CP-08}
Colomo F., Pronko A.G., The limit shape of large alternating sign matrices,
 \href{https://doi.org/10.1137/080730639}{\textit{SIAM~J. Discrete Math.}} \textbf{24} (2010), 1558--1571,
 \href{https://arxiv.org/abs/0803.2697}{arXiv:0803.2697}.

\bibitem{CP-12}
Colomo F., Pronko A.G., An approach for calculating correlation functions in
 the six-vertex model with domain wall boundary conditions, \href{https://doi.org/10.1007/s11232-012-0061-2}{\textit{Theoret.
 and Math. Phys.}} \textbf{171} (2012), 641--654, \href{https://arxiv.org/abs/1111.4353}{arXiv:1111.4353}.

\bibitem{dB-58}
de~Bruijn N.G., Asymptotic methods in analysis, \textit{Bibliotheca
 Mathematica}, Vol.~4, North-Holland Publishing Co., Amsterdam, 1958.

\bibitem{E-81}
Erd\'elyi A., Magnus W., Oberhettinger F., Tricomi F.G., Higher transcendental
 functions. {V}ol.~{I}, Robert E. Krieger Publishing Co., Inc., Melbourne,
 Fla., 1981.

\bibitem{EGSV-11a}
Escobedo J., Gromov N., Sever A., Vieira P., Tailoring three-point functions
 and integrability, \href{https://doi.org/10.1007/JHEP09(2011)028}{\textit{J.~High Energy Phys.}} \textbf{2011} (2011), no.~9,
 028, 50~pages, \href{https://arxiv.org/abs/1012.2475}{arXiv:1012.2475}.

\bibitem{EGSV-11b}
Escobedo J., Gromov N., Sever A., Vieira P., Tailoring three-point functions
 and integrability~{II}. {W}eak/strong coupling match, \href{https://doi.org/10.1007/JHEP09(2011)029}{\textit{J.~High Energy
 Phys.}} \textbf{2011} (2011), no.~9, 029, 35~pages, \href{https://arxiv.org/abs/1104.5501}{arXiv:1104.5501}.

\bibitem{F-77}
Fedoryuk M.V., The saddle-point method, Nauka, Moscow, 1977.

\bibitem{F-12}
Foda O., {$\mathcal N=4$} {SYM} structure constants as determinants,
 \href{https://doi.org/10.1007/JHEP03(2012)096}{\textit{J.~High Energy Phys.}} \textbf{2012} (2012), no.~3, 096, 29~pages,
 \href{https://arxiv.org/abs/1111.4663}{arXiv:1111.4663}.

\bibitem{FW-12}
Foda O., Wheeler M., Partial domain wall partition functions, \href{https://doi.org/10.1007/JHEP07(2012)186}{\textit{J.~High
 Energy Phys.}} \textbf{2012} (2012), no.~7, 186, 36~pages, \href{https://arxiv.org/abs/1205.4400}{arXiv:1205.4400}.

\bibitem{G-83}
Gaudin M., La fonction d'onde de {B}ethe, \textit{Collection du Commissariat \`a
 l'\'Energie Atomique: S\'erie Scientifique.}, Masson, Paris, 1983.

\bibitem{GSV-12}
Gromov N., Sever A., Vieira P., Tailoring three-point functions and
 integrability~{III}. {C}lassical tunneling, \href{https://doi.org/10.1007/JHEP07(2012)044}{\textit{J.~High Energy Phys.}}
 \textbf{2012} (2012), no.~7, 044, 31~pages, \href{https://arxiv.org/abs/1111.2349}{arXiv:1111.2349}.

\bibitem{GS-92}
Gwa L.-H., Spohn H., Six-vertex model, roughened surfaces, and an asymmetric
 spin {H}amiltonian, \href{https://doi.org/10.1103/PhysRevLett.68.725}{\textit{Phys. Rev. Lett.}} \textbf{68} (1992), 725--728.

\bibitem{I-87}
Izergin A.G., Partition function of the six-vertex model in the finite volume,
 \textit{Sov. Phys. Dokl.} \textbf{32} (1987), 878--879.

\bibitem{ICK-92}
Izergin A.G., Coker D.A., Korepin V.E., Determinant formula for the six-vertex
 model, \href{https://doi.org/10.1088/0305-4470/25/16/010}{\textit{J.~Phys.~A: Math. Gen.}} \textbf{25} (1992), 4315--4334.

\bibitem{IKR-87}
Izergin A.G., Korepin V.E., Reshetikhin N.Yu., Correlation functions in a
 one-dimensional {B}ose gas, \href{https://doi.org/10.1088/0305-4470/20/14/022}{\textit{J.~Phys.~A: Math. Gen.}} \textbf{20}
 (1987), 4799--4822.

\bibitem{JKKS-16}
Jiang Y., Komatsu S., Kostov I., Serban D., The hexagon in the mirror: the
 three-point function in the {S}o{V} representation, \href{https://doi.org/10.1088/1751-8113/49/17/174007}{\textit{J.~Phys.~A: Math.
 Theor.}} \textbf{49} (2016), 174007, 31~pages, \href{https://arxiv.org/abs/1506.09088}{arXiv:1506.09088}.

\bibitem{KP-20}
Kapitonov V.S., Pronko A.G., Six-vertex model as a {G}rassmann integral,
 one-point function, and the arctic ellipse, \textit{Zap. Nauchn. Semin.
 POMI} \textbf{494} (2020), 168--218.

\bibitem{KP-16}
Kitaev A.V., Pronko A.G., Emptiness formation probability of the six-vertex
 model and the sixth {P}ainlev\'e equation, \href{https://doi.org/10.1007/s00220-016-2636-5}{\textit{Comm. Math. Phys.}}
 \textbf{345} (2016), 305--354, \href{https://arxiv.org/abs/1505.00032}{arXiv:1505.00032}.

\bibitem{KKMST-09}
Kitanine N., Kozlowski K.K., Maillet J.M., Slavnov N.A., Terras V., Algebraic
 {B}ethe ansatz approach to the asymptotic behavior of correlation functions,
 \href{https://doi.org/10.1088/1742-5468/2009/04/p04003}{\textit{J.~Stat. Mech. Theory Exp.}} \textbf{2009} (2009), P04003, 66~pages, \href{https://arxiv.org/abs/0808.0227}{arXiv:0808.0227}.

\bibitem{KMST-02}
Kitanine N., Maillet J.M., Slavnov N.A., Terras V., Spin-spin correlation
 functions of the {$XXZ\text{-}{\frac{1}{2}}$} {H}eisenberg chain in a magnetic
 field, \href{https://doi.org/10.1016/S0550-3213(02)00583-7}{\textit{Nuclear Phys.~B}} \textbf{641} (2002), 487--518,
 \href{https://arxiv.org/abs/hep-th/0201045}{arXiv:hep-th/0201045}.

\bibitem{K-82}
Korepin V.E., Calculation of norms of {B}ethe wave functions, \href{https://doi.org/10.1007/BF01212176}{\textit{Comm.
 Math. Phys.}} \textbf{86} (1982), 391--418.

\bibitem{KBI-93}
Korepin V.E., Bogoliubov N.M., Izergin A.G., Quantum inverse scattering method
 and correlation functions, \textit{Cambridge Monographs on Mathematical Physics},
 \href{https://doi.org/10.1017/CBO9780511628832}{Cambridge University Press}, Cambridge, 1993.

\bibitem{KZj-00}
Korepin V.E., Zinn-Justin P., Thermodynamic limit of the six-vertex model with
 domain wall boundary conditions, \href{https://doi.org/10.1088/0305-4470/33/40/304}{\textit{J.~Phys.~A: Math. Gen.}} \textbf{33}
 (2000), 7053--7066, \href{https://arxiv.org/abs/cond-mat/0004250}{arXiv:cond-mat/0004250}.

\bibitem{K-12a}
Kostov I., Classical limit of the three-point function of {$N=4$}
 supersymmetric {Y}ang--{M}ills theory from integrability, \href{https://doi.org/10.1103/PhysRevLett.108.261604}{\textit{Phys. Rev.
 Lett.}} \textbf{108} (2012), 261604, 5~pages, \href{https://arxiv.org/abs/1203.6180}{arXiv:1203.6180}.

\bibitem{K-12b}
Kostov I., Three-point function of semiclassical states at weak coupling,
 \href{https://doi.org/10.1088/1751-8113/45/49/494018}{\textit{J.~Phys.~A: Math. Theor.}} \textbf{45} (2012), 494018, 27~pages,
 \href{https://arxiv.org/abs/1205.4412}{arXiv:1205.4412}.

\bibitem{KS-82}
Kulish A.G., Sklyanin E.K., Solutions of the {Y}ang--{B}axter equation,
 \href{https://doi.org/10.1007/BF01091463}{\textit{J.~Sov. Math.}} \textbf{19} (1982), 1596--1620.

\bibitem{Ku-96}
Kuperberg G., Another proof of the alternating-sign matrix conjecture,
 \href{https://doi.org/10.1155/S1073792896000128}{\textit{Int. Math. Res. Not.}} \textbf{1996} (1996), 139--150,
 \href{https://arxiv.org/abs/math.CO/9712207}{arXiv:math.CO/9712207}.

\bibitem{LKRV-18}
Lyberg I., Korepin V., Ribeiro G.A.P., Viti J., Phase separation in the
 six-vertex model with a variety of boundary conditions, \href{https://doi.org/10.1063/1.5018324}{\textit{J.~Math.
 Phys.}} \textbf{59} (2018), 053301, 15~pages, \href{https://arxiv.org/abs/1711.07905}{arXiv:1711.07905}.

\bibitem{MP-19}
Minin M.D., Pronko A.G., Boundary polarization of the rational six-vertex model
 on a semi-infinite lattice, \href{https://doi.org/10.1007/s10958-021-05501-4}{\textit{J.~Math. Sci.}} \textbf{257} (2021),
 537--550.

\bibitem{PP-19}
Pronko A.G., Pronko G.P., Off-shell {B}ethe states and the six-vertex model,
 \href{https://doi.org/10.1007/s10958-019-04511-7}{\textit{J.~Math. Sci.}} \textbf{242} (2019), 742--752.

\bibitem{SZ-04}
Sylju{\aa}sen O.F., Zvonarev M.B., Directed-loop {M}onte {C}arlo simulations of
 vertex models, \href{https://doi.org/10.1103/PhysRevE.70.016118}{\textit{Phys. Rev.~E}} \textbf{70} (2004), 016118, 8~pages,
 \href{https://arxiv.org/abs/cond-mat/0401491}{arXiv:cond-mat/0401491}.

\bibitem{TF-79}
Takhtajan L.A., Faddeev L.D., The quantum method for the inverse problem and
 the {H}eisenberg {$XYZ$} model, \href{https://doi.org/10.1070/RM1979v034n05ABEH003909}{\textit{Russian Math. Surveys}} \textbf{34}
 (1979), no.~5, 11--68.

\bibitem{W-11}
Wheeler M., An {I}zergin--{K}orepin procedure for calculating scalar products
 in the six-vertex model, \href{https://doi.org/10.1016/j.nuclphysb.2011.07.006}{\textit{Nuclear Phys.~B}} \textbf{852} (2011),
 468--507, \href{https://arxiv.org/abs/1104.2113}{arXiv:1104.2113}.

\bibitem{Ze-96}
Zeilberger D., Proof of the refined alternating sign matrix conjecture,
 \textit{New York~J. Math.} \textbf{2} (1996), 59--68,
 \href{https://arxiv.org/abs/math.CO/9606224}{arXiv:math.CO/9606224}.

\end{thebibliography}
\end{document}